\documentclass[journal=jacsat,manuscript=article]{achemso}
\setkeys{acs}{articletitle = true, chaptertitle = true}

\usepackage[version=3]{mhchem}
\usepackage{url}
\usepackage{booktabs}
\usepackage{graphicx}
\usepackage{spverbatim}
\usepackage{fancyvrb}    
\usepackage{mdframed}    
\usepackage{xcolor}      
\usepackage{atbegshi}
\definecolor{lightgray}{gray}{0.95} 
\newmdenv[
  backgroundcolor=lightgray,
  linecolor=lightgray,     
  skipabove=10pt,
  skipbelow=10pt,
  innerleftmargin=10pt,
  innerrightmargin=10pt,
  innertopmargin=8pt,
  innerbottommargin=8pt,
  nobreak=true
]{grayverb}
\newcommand{\commandboxneedspace}{\par\needspace{8\baselineskip}}
\newcounter{appendixsection}
\renewcommand{\theappendixsection}{\Alph{appendixsection}}
\newcommand{\numberedappendix}[1]{%
  \refstepcounter{appendixsection}%
  \section*{Appendix \theappendixsection: #1}%
}

\usepackage{hyperref}
\hypersetup{
  colorlinks=true,
  linkcolor=blue,      
  citecolor=blue,      
  urlcolor=blue,       
  breaklinks=true
}
\AtBeginShipoutNext{\AtBeginShipoutDiscard\global\setcounter{page}{0}}

\author{Fengbo Yuan}
\affiliation{Department of Physics, University of Alabama at Birmingham, Birmingham, Alabama, 35205, USA}
\author{Xin Zhong}
\affiliation{Institut für Geologische Wissenschaften, Freie Universität Berlin, 12249 Berlin, Germany}
\author{Donghao Zheng}
\affiliation{Department of Geosciences, Princeton University, 11 Ivy Lane, Princeton, NJ 08550, USA}
\author{Jinzhe Zeng}
\affiliation{School of Artificial Intelligence and Data Science, University of Science and Technology of China, Hefei 230026, P.R.~China}
\alsoaffiliation{Suzhou Institute for Advanced Research, University of Science and Technology of China, Suzhou 215123, P.R.~China}
\alsoaffiliation{Suzhou Big Data \& AI Research and Engineering Center, Suzhou 215123, P.R.~China}
\author{Linfeng Zhang}
\affiliation{AI for Science Institute, Beijing 100080, P. R. China}
\alsoaffiliation{DP Technology, Beijing 100080, P. R. China}
\author{Han Wang}
\affiliation{
National Key Laboratory of Computational Physics, Institute of Applied Physics and Computational Mathematics, Fenghao East Road 2, Beijing 100094, P. R. China
}
\alsoaffiliation{HEDPS, CAPT, College of Engineering, Peking University, Beijing 100871, P. R. China}
\author{Yifan Li}
\affiliation{
Department of Chemistry, Princeton University, Princeton, NJ 08544, USA
}
\email{yifanl0716@gmail.com}
\title{dpti: An Automated Thermodynamic Integration Workflow for Phase Diagram Calculations with Machine Learning Interatomic Potentials}

\begin{document}

\begin{tocentry}
\includegraphics[width=\linewidth]{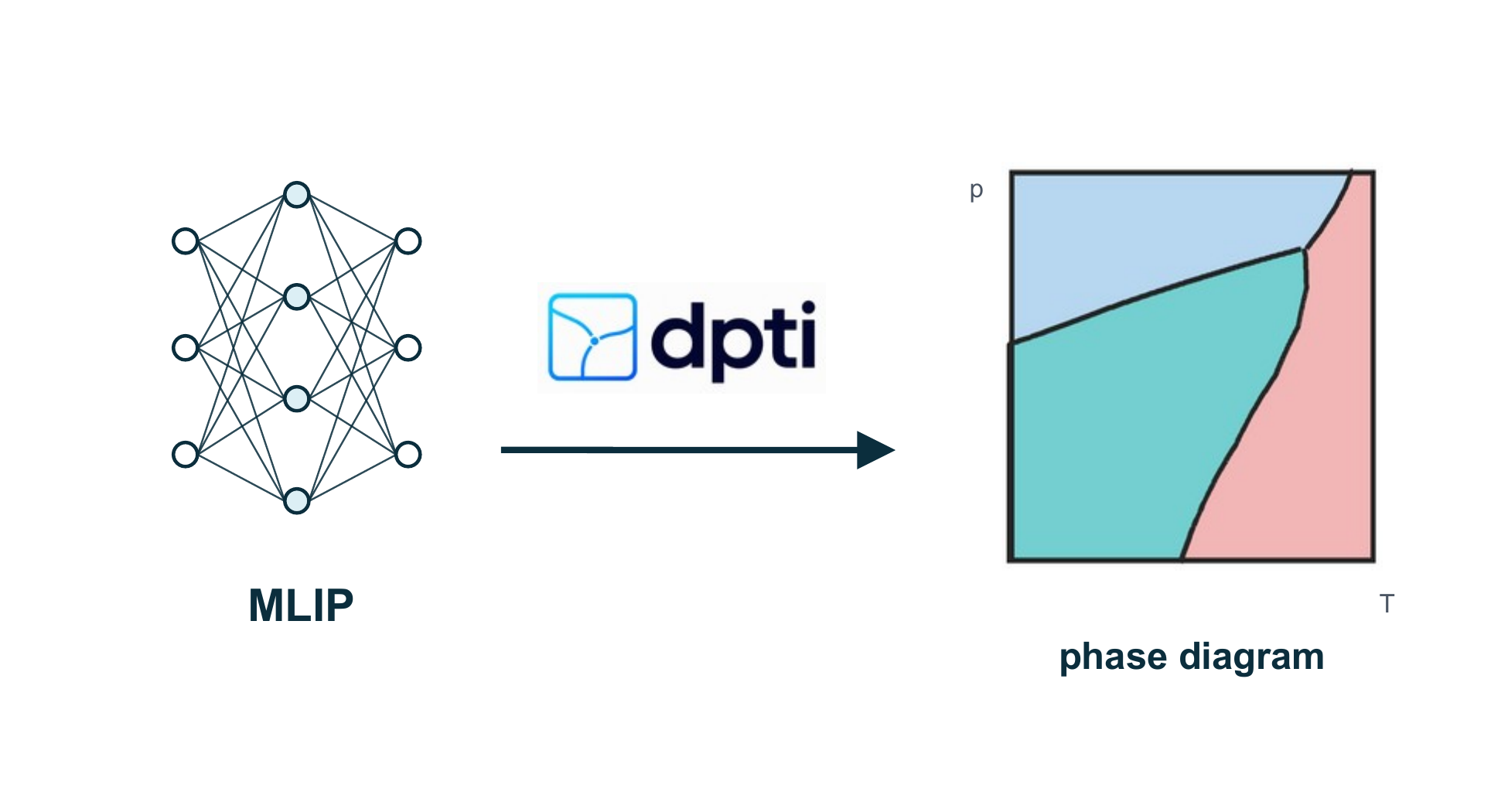}
\end{tocentry}

\begin{abstract}
  Thermodynamic integration (TI) is a widely used approach for computing free energies and phase diagrams. However, TI calculations driven by machine learning interatomic potentials (MLIPs) remain technically challenging because they require careful design of reversible integration paths and many closely related molecular dynamics (MD) tasks for each phase and state point. To address these challenges, we present \texttt{dpti}, an open-source Python package that automates TI workflows for phase diagram calculations with MLIPs. \texttt{dpti} connects reference systems with analytically known free energies to MLIP-described atomic and molecular solids and liquids through reversible integration paths. Given JSON input files, \texttt{dpti} generates and runs the required MD tasks, computes free energy contributions, estimates errors, and propagates coexistence points into phase boundaries. We demonstrate the usage of \texttt{dpti} with two examples driven by Deep Potential models: a silica phase diagram involving $\beta$-quartz, coesite, and melt, and the ice Ih--liquid water phase boundary. \texttt{dpti} provides a useful tool for automated phase diagram calculations of materials modeled by MLIPs.
\end{abstract}

\section{Introduction}
Phase diagrams are central to materials science~\cite{chew_phase_2023}. Predicting phase diagrams from atomistic simulations requires accurate free energy differences between competing phases over a range of temperatures and pressures. Among free energy methods, thermodynamic integration (TI), which connects systems or state points through a reversible integration path, is one of the most widely used approaches. For solids, Frenkel and Ladd~\cite{frenkel_new_1984} established the Einstein crystal method for absolute free energy calculations, which was later complemented by finite-size corrections by Polson et al.~\cite{polson_finite-size_2000} and by the Einstein molecule approach of Vega et al.~\cite{vega_revisiting_2007,vega_determination_2008}. Sugino and Car~\cite{sugino_ab_1995} extended the TI framework to \textit{ab initio} molecular dynamics (AIMD) simulations, enabling \textit{ab initio} phase diagram calculations for materials. Over the past decades, TI has been successfully applied in molecular dynamics (MD) studies of phase diagrams for a broad range of materials, including iron~\cite{alfe_melting_1999}, carbon~\cite{ghiringhelli_modeling_2005,wang_carbon_2005}, silicon~\cite{sugino_ab_1995,kaczmarski_phase_2005}, and silica~\cite{saika-voivod_phase_2004,ford_further_2007}.

Machine learning interatomic potentials (MLIPs)~\cite{zhang_deep_2018,zhang_end,wang_deepmd-kit_2018,zeng_deepmd-kit_2023,zeng_deepmd-kit_2025} have enabled efficient simulations of materials with \textit{ab initio} accuracy and have become increasingly powerful tools for phase diagram prediction. However, TI calculations with MLIPs remain technically demanding. First, constructing a reversible integration path from a reference system with an analytically known free energy to a target system described by an MLIP is nontrivial, especially for complex materials such as molecular solids and liquids. Second, TI calculations require many closely related MD tasks for each phase, making workflow management and reproducibility difficult when the tasks are prepared and post-processed manually. Third, the predicted phase boundary can be sensitive to small free energy errors, so both statistical uncertainty and numerical integration error must be carefully monitored. These challenges motivate a systematic and automated workflow for setting up TI calculations and analyzing the resulting free energy data.

Several automated workflow tools have been developed for free energy methods, including TI. de Koning and co-workers developed nonequilibrium free energy methods based on adiabatic switching and reversible scaling, including LAMMPS implementations for solids and fluids~\cite{de_koning_einstein_1996,freitas_nonequilibrium_2016,paula_leite_nonequilibrium_2019,plimpton_fast_1995,thompson_lammps_2022}. Building on this LAMMPS-based nonequilibrium free energy framework, Menon et al. developed CALPHY~\cite{menon_automated_2021} to automate calculations of absolute free energies, phase boundaries, and alchemical and upscaling free energies. Other free energy tools address related but distinct contexts: PLUMED provides a general interface for enhanced sampling methods~\cite{bonomi_plumed_2009,tribello_plumed_2014}, whereas OpenFE focuses on alchemical free energy workflows, particularly ligand binding and hydration free energies~\cite{alibay_httpsgithubcomopenfreeenergyopenfe_2025}. In comparison, our goal is to automate a workflow specifically for MLIP-based phase-diagram calculations using equilibrium TI.

Here we present \texttt{dpti}, an open-source Python package for automating TI workflows for phase diagram calculations with MLIPs. We call it \texttt{dpti} because it was motivated by phase diagram calculations with the Deep Potential (DP) model~\cite{zhang_phase_2021}, but the workflow is general and can be applied to other MLIPs. \texttt{dpti} currently supports atomic solids, atomic liquids, and water as a representative molecular system. More complex molecular systems require careful treatment of molecular topology to design reversible integration paths and are not yet supported. The package connects reference interactions to MLIPs through reversible integration paths with intermediate states tailored to different classes of systems. It generates the large number of related LAMMPS input files~\cite{plimpton_fast_1995, thompson_lammps_2022} required for TI calculations and manages their execution on HPC clusters. It also provides post-processing tools for statistical error propagation, numerical integration-error estimation, and adaptive refinement of integration grids. In this way, \texttt{dpti} turns the repetitive tasks of phase diagram calculations into reproducible and automated workflows.

The remainder of this paper is organized as follows. We first describe the theoretical framework underlying the \texttt{dpti} workflow, including reference-state construction, integration paths, consistency checks, and error estimation. We then demonstrate \texttt{dpti} through two representative examples: the coesite--$\beta$-quartz--melt phase boundaries in silica and the ice Ih--liquid water phase boundary. Finally, we discuss the current scope of \texttt{dpti}, its previous applications, and remaining challenges for extending MLIP-based free energy calculations to more complex molecular systems.

\section{Theory}
In this section, we briefly describe the theoretical framework for TI-based phase diagram calculations implemented in \texttt{dpti}. Readers are referred to the review by Vega et al.~\cite{vega_determination_2008} for a comprehensive introduction to TI methods for phase diagram calculations.

\subsection{General Workflow for Phase Diagram Calculations with \texttt{dpti}}
\begin{figure}[!ht]
\centering
\includegraphics[width=\textwidth]{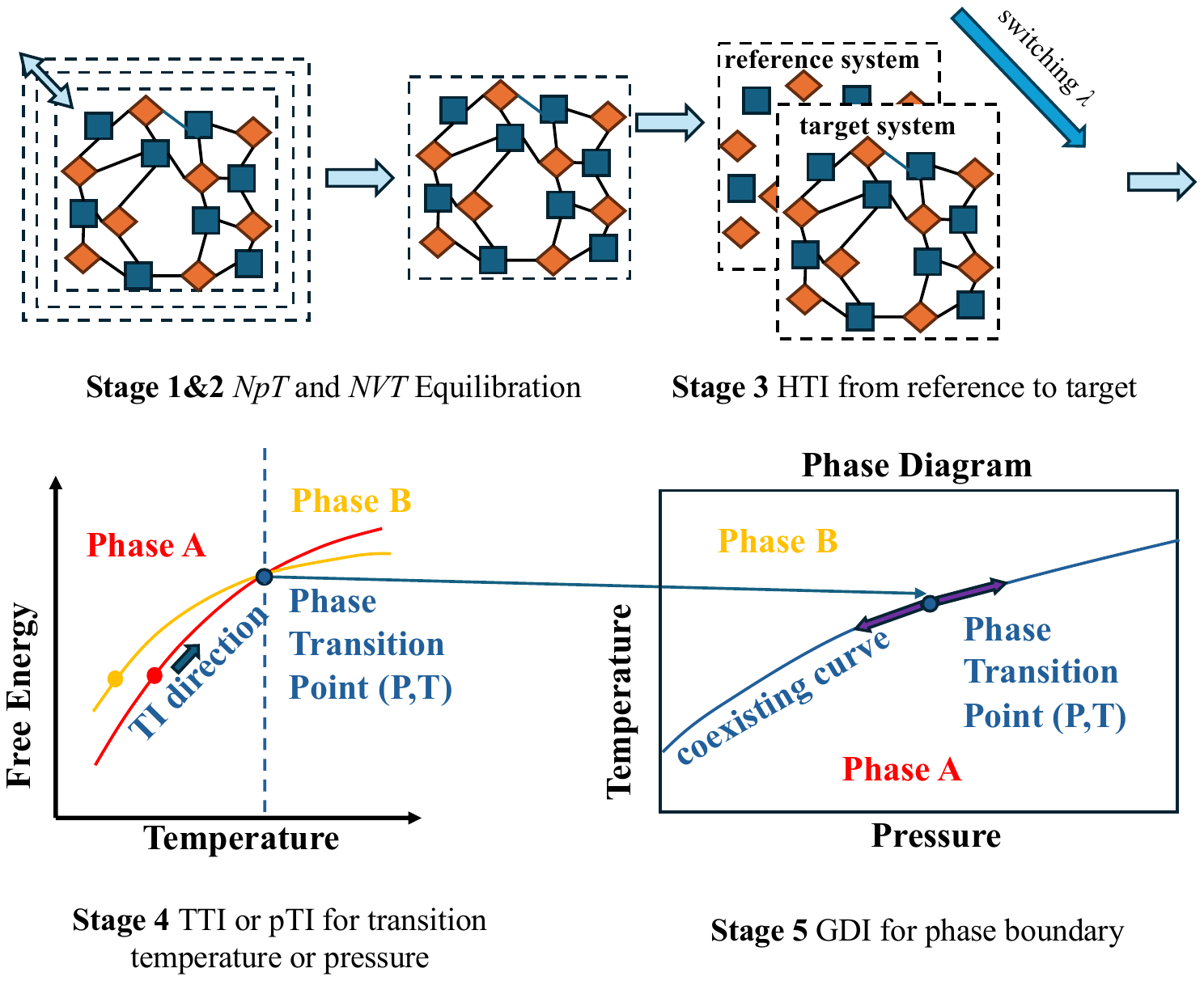}
\caption{Schematic workflow of \texttt{dpti} for phase diagram calculations. Stages 1 and 2 prepare the target phase by \(NpT\) and \(NVT\) equilibration. Stage 3 performs Hamiltonian thermodynamic integration from the reference to target state. Stage 4 propagates the Gibbs free energy with TTI or pTI to locate coexistence points. Stage 5 propagates the coexistence point into a phase boundary by Gibbs-Duhem integration.}
\label{fig:workflow}
\end{figure}

The standard \texttt{dpti} workflow for phase diagram calculations consists of five stages, as summarized in Fig.~\ref{fig:workflow}. The first four stages can be run independently for each phase, because they give the Gibbs free energy of each phase within a range of temperatures or pressures. The fifth stage should be run for pairs of phases to trace their coexistence line.

First, an $NpT$ simulation is performed to determine the average density or cell parameters of the target phase at a chosen state point $(p_0,T_0)$. In the current \texttt{dpti} workflow, Hamiltonian thermodynamic integration is then performed in the $NVT$ ensemble using this fixed cell, and the resulting Helmholtz free energy is converted to a Gibbs free energy according to Eq.~\eqref{eq:gibbs_from_helmholtz}. This $NVT$-based approach is common practice in TI calculations, although it is possible to perform the entire TI procedure directly in the $NpT$ ensemble~\cite{de_witte_novel_2026}.

Second, an $NVT$ simulation equilibrates the atomic configuration at the average cell parameters to generate an initial configuration for subsequent free energy calculations.

Third, Hamiltonian thermodynamic integration (HTI) is used to compute the absolute Helmholtz free energy of the phase relative to a reference state whose free energy can be computed analytically. One interpolates between the reference Hamiltonian $H_0$ and the target Hamiltonian $H_1$ using a coupling parameter $\lambda$:
\begin{equation}
H(\lambda) = (1-\lambda)H_0 + \lambda H_1.
\end{equation}
Here $H_0$ and $H_1$ have the potential energies $U_0$ and $U_1$, respectively. Similarly, the interpolated potential energy is $U(\lambda)=(1-\lambda)U_0 + \lambda U_1$. 
The corresponding Helmholtz free energy difference is
\begin{equation}\label{eq:hti}
A_1 - A_0
=
\int_0^1 \left\langle \frac{\partial H(\lambda)}{\partial \lambda} \right\rangle_{\lambda} \, \mathrm{d}\lambda
=
\int_0^1 \left\langle \frac{\partial U(\lambda)}{\partial \lambda} \right\rangle_{\lambda} \, \mathrm{d}\lambda
=
\int_0^1 \left\langle U_1-U_0 \right\rangle_{\lambda} \, \mathrm{d}\lambda,
\end{equation}
where $\langle \cdot \rangle_{\lambda}$ denotes the canonical ensemble average at the interpolated Hamiltonian $H(\lambda)$~\cite{frenkel_new_1984,vega_determination_2008}.

Eq.~\eqref{eq:hti} is only valid when the TI path from the reference to the target state is reversible. Here, a reversible path means a sequence of intermediate Hamiltonians for which each state remains in equilibrium, so that the forward and reverse integrations would give the same free energy difference. This condition can fail if a first-order phase transition occurs along the path, or if kinetic trapping or glassy dynamics makes the sampled state history dependent. In addition, the ensemble average at each \(\lambda\) must be sufficiently converged for the integration to yield the correct free energy difference.

After obtaining the Helmholtz free energy difference, the corresponding Gibbs free energy is obtained by adding the $pV$ contribution at the same state point,
\begin{equation}\label{eq:gibbs_from_helmholtz}
G(p,T) = A(\langle V \rangle_{p,T},T) + p\langle V \rangle_{p,T},
\end{equation}
where $\langle V \rangle_{p,T}$ is the average volume at pressure $p$ and temperature $T$.

Fourth, temperature thermodynamic integration (TTI) and pressure thermodynamic integration (pTI) propagate the Gibbs free energy from the reference state point to other temperatures and pressures. Given the Gibbs free energy at $(p_0,T_0)$, TTI propagates it along an isobar via the Gibbs-Helmholtz equation. Defining the enthalpy as $\mathcal{H}=K+U+pV$, where \(K\) and \(U\) are the kinetic and potential energies, one obtains~\cite{vega_determination_2008,zhang_phase_2021}
\begin{equation}\label{eq:tti}
\frac{G(p_0,T_1)}{k_{\mathrm B} T_1}
-
\frac{G(p_0,T_0)}{k_{\mathrm B} T_0}
=
-\int_{T_0}^{T_1}
\frac{\langle \mathcal{H} \rangle_{p_0,T}}{k_{\mathrm B} T^2}
\, \mathrm{d}T.
\end{equation}
Similarly, pTI propagates the free energy along an isotherm:
\begin{equation}\label{eq:pti}
G(p_1,T_0)
-
G(p_0,T_0)
=
\int_{p_0}^{p_1}
\langle V \rangle_{p,T_0}
\,\mathrm{d}p.
\end{equation}
In \texttt{dpti}, these temperature and pressure integrals are evaluated numerically from a discrete set of $NpT$ simulations. In practice, the choice between TTI and pTI depends on the local slope of the phase boundary. When $\partial p/\partial T$ is close to zero, the coexistence line is nearly horizontal in the $p$--$T$ plane, and pTI is usually the more convenient route, as it locates the phase boundary from free-energy crossings along isotherms. Otherwise, TTI is typically preferred because it tracks free-energy crossings along isobars. As in HTI, no first-order phase transition should occur along TTI and pTI paths so that the integrand remains smooth and the integrations in Eqs.~\eqref{eq:tti} and \eqref{eq:pti} yield the correct free energy differences.

Fifth, once a coexistence point between two phases has been located from the crossing of their Gibbs free energies, Gibbs-Duhem integration (GDI) is used to trace the phase boundary starting from that coexistence point. For two coexisting phases $\alpha$ and $\beta$, $G_{\alpha}(p,T)=G_{\beta}(p,T)$ along the coexistence line, and thus
\begin{equation}
\frac{\mathrm{d}p}{\mathrm{d}T}
=
\frac{1}{T}
\frac{\langle H \rangle_{p,T}^{\alpha}-\langle H \rangle_{p,T}^{\beta}}
{\langle V \rangle_{p,T}^{\alpha}-\langle V \rangle_{p,T}^{\beta}},
\label{eq:gdi}
\end{equation}
which is the Clausius-Clapeyron equation written in terms of ensemble-averaged enthalpy and volume differences between the two phases~\cite{kofke_direct_1993,zhang_phase_2021}. Therefore, a single coexistence point, together with separate $NpT$ simulations for the two coexisting phases, is sufficient to propagate a phase boundary over a finite pressure or temperature range~\cite{frenkel_new_1984,vega_determination_2008,kofke_direct_1993,zhang_phase_2021}.

\subsection{Reference States for HTI}
A reference state for HTI should have an analytically known free energy and be connected to the target state through a reversible integration path. 
\subsubsection{Atomic Solids and Liquids}
For atomic solids, \texttt{dpti} uses Einstein-type reference states in which atoms are harmonically restrained to reference positions,
\begin{equation}
H_0
=
\sum_s\sum_{i\in s}\frac{p_i^2}{2m_s}
+
\sum_s\sum_{i\in s}\frac{1}{2}k_s(\mathbf{r}_i-\mathbf{r}_{i,0})^2,
\end{equation}
where $s$ indexes atomic species, $i\in s$ labels atoms of species $s$, $\mathbf{r}_{i,0}$ are the equilibrium lattice positions, and $k_s$ is the harmonic spring constant for species $s$. The free energy of this unconstrained Einstein reference can be evaluated analytically,
\begin{equation}\label{eq:einstein_reference}
A_0^{\mathrm E}
=
3k_{\mathrm B}T
\sum_s N_s
\left(
\ln \Lambda_s
+
\ln \Gamma_s
\right),
\end{equation}
where $N_s$ is the number of atoms of species $s$, $\Lambda_s=h/\sqrt{2\pi m_s k_{\mathrm B}T}$ is the thermal wavelength, and $\Gamma_s=\sqrt{k_s/(2\pi k_{\mathrm B}T)}$ is the inverse configurational length associated with the species-dependent spring constant. In practice, \(k_s\) should be chosen so that the Einstein-crystal mean-square displacement, \(N_s^{-1}\sum_{i\in s}\langle |\mathbf{r}_i-\mathbf{r}_{i,0}|^2\rangle=3k_{\mathrm B}T/k_s\), is comparable to that of the target solid at the same temperature. The species-dependent estimates \(k_s\) can be converted to a common mass-normalized value as \(\kappa=(\sum_s x_s m_s/k_s)^{-1}\), where \(x_s\) is the atomic fraction of species \(s\). This estimate can then be checked by repeating the HTI calculation with nearby spring constants.

In practical HTI calculations, however, directly connecting the unconstrained Einstein crystal to a periodic target solid leads to an ill-behaved integration path because the target solid has a translational degree of freedom. When the harmonic springs vanish near the target state, the crystal can drift as a whole relative to the reference lattice, producing a sharp peak in the integrand~\cite{frenkel_new_1984}. This issue is usually handled by one of two approaches, Frenkel's Einstein crystal or Vega's Einstein molecule. In the Frenkel approach, the center of mass (CM) is fixed throughout the HTI path~\cite{frenkel_new_1984}. In the Vega approach, one reference particle is fixed~\cite{vega_revisiting_2007}. In both cases, the integration is performed between a constrained reference state and the target solid under the same constraint. The free energy of the unconstrained target solid is then recovered by adding the corresponding analytical correction. The effective reference free energies used by \texttt{dpti} for the Frenkel and Vega formulations are summarized in Appendices~\ref{app:cm_correction} and \ref{app:vega_correction}, respectively.

For atomic liquids, \texttt{dpti} starts from an ideal-gas reference. The ideal-gas reference free energy is
\begin{equation}
A_0^{\mathrm{id}}
=
k_{\mathrm B}T
\sum_s
\left[
N_s\ln(\rho_s\Lambda_s^3)
-N_s
+\frac{1}{2}\ln(2\pi N_s)
\right],
\end{equation}
where $\rho_s=N_s/V$. Removing the total momentum in liquid simulations does not change the configurational free energy and therefore introduces no additional free energy correction.

\subsubsection{Water}
The reference state for water requires special consideration because water contains complex intramolecular and intermolecular interactions.

For ice phases, \texttt{dpti} uses the same treatment as for atomic solids. An additional contribution to the free energy, $-TS_{\mathrm{conf}}$, arises from proton disorder. The entropy of fully disordered phases such as ice Ih, Ic, IV, VI, VII, and XII is accounted for using Pauling's approximation~\cite{pauling_structure_1935}, $S_{\mathrm{conf}}/N_{\mathrm{H_2O}} \approx k_{\mathrm B}\ln 1.5$, where $N_{\mathrm{H_2O}}$ is the number of water molecules, whereas partially disordered phases such as ice III and V use the combinatorial entropy estimates of Macdowell et al.~\cite{macdowell_combinatorial_2004}. The corresponding reference free energy is given in Appendix~\ref{app:ice_reference}.


For liquid water, a specialized molecular reference state is used. The reference is an ideal gas of noninteracting water molecules in which each oxygen atom is connected to two hydrogen atoms by harmonic O--H springs, while the H--O--H angle is not constrained. The corresponding reference free energy is given in Appendix~\ref{app:water_reference}. Water is a representative molecular system for constructing HTI reference states. The strategy of building a molecular reference by retaining selected intramolecular coordinates can be transferred to other molecular materials with appropriate choices of molecular constraints and reference potentials.

\subsection{HTI Paths from Reference to Target States}
During HTI, the system is transformed from the reference state to the target state by gradually switching to the target potential. The choice of the HTI path is crucial for ensuring the reversibility of the transformation and the validity of the MLIP. 
First, first-order phase transitions should be avoided along the HTI path so that the path is reversible and the integrand is continuous.
Second, MLIPs are usually trained on configurations sampled near the target state and may become unreliable if the intermediate configurations along the HTI path deviate too much from the target state. Therefore, the HTI path should be designed to keep the system close to the target state and avoid sharp changes in the integrand. 

\subsubsection{Atomic Systems}
For atomic solids, \texttt{dpti} supports one-step, two-step, and three-step paths. The one-step path switches on the target potential while switching off the harmonic springs simultaneously,
\begin{equation}
U(\lambda)=(1-\lambda)U_{\mathrm{spring}}+\lambda U_{\mathrm{target}}.
\end{equation}
The two-step path first switches on the target potential with the springs retained and then removes the springs,
\begin{align}
U_1(\lambda)&=U_{\mathrm{spring}}+\lambda U_{\mathrm{target}},\\
U_2(\lambda)&=U_{\mathrm{target}}+(1-\lambda)U_{\mathrm{spring}}.
\end{align}
The three-step path introduces a soft-core Lennard-Jones (LJ) interaction before the target potential is turned on and removes the LJ interaction together with the springs in the final step,
\begin{align}
U_1(\lambda)&=U_{\mathrm{spring}}+\lambda U_{\mathrm{LJ}},\\
U_2(\lambda)&=U_{\mathrm{spring}}+U_{\mathrm{LJ}}+\lambda U_{\mathrm{target}},\\
U_3(\lambda)&=U_{\mathrm{target}}+(1-\lambda)(U_{\mathrm{spring}}+U_{\mathrm{LJ}}).
\end{align}
Here $U_{\mathrm{LJ}}$ denotes a pairwise soft-core LJ potential. For a pair distance $r<r_c$, its pair contribution is
\begin{equation}
u_{\mathrm{LJ}}^{\mathrm{soft}}(r)
=
\eta^n 4\epsilon
\left[
\frac{1}{\left[\alpha(1-\eta)^2+(r/\sigma)^6\right]^2}
-
\frac{1}{\alpha(1-\eta)^2+(r/\sigma)^6}
\right],
\end{equation}
and it is set to zero for $r\ge r_c$. The soft-core LJ interaction is specified by six parameters: $n$ controls the power-law scaling with the activation parameter $\eta$, $\alpha$ controls the softness of the repulsive core, $\epsilon$ and $\sigma$ are the usual LJ energy and length scales, and $r_c$ is the cutoff radius.

For atomic liquids, HTI transforms an ideal-gas reference at the target liquid volume into the target liquid phase using a three-step path. The auxiliary soft-core LJ interaction regularizes close contacts in the dense ideal-gas reference and generates an interacting fluid-like intermediate before the target potential is switched on. The three-step path switches on the soft-core LJ interaction, switches on the target potential, and finally switches off the soft-core LJ interaction:
\begin{align}
U_1(\lambda)&=\lambda U_{\mathrm{LJ}},\\
U_2(\lambda)&=U_{\mathrm{LJ}}+\lambda U_{\mathrm{target}},\\
U_3(\lambda)&=U_{\mathrm{target}}+(1-\lambda)U_{\mathrm{LJ}}.
\end{align}

\subsubsection{Water}
For ice, the HTI path is the same as that for atomic solids. The three-step path is recommended for ice phases. 

For liquid water, the HTI path introduces an angular restraint together with the soft-core LJ interaction, turns on the target potential, and removes the auxiliary LJ and bond-angle reference terms~\cite{zhang_phase_2021}. In \texttt{dpti}, these steps correspond to an angle-on step, a target-on step, and a bond-angle-off step:
\begin{align}
U_1(\lambda)&=U_{\mathrm{bond}}+\lambda(U_{\mathrm{angle}}+U_{\mathrm{LJ}}),\\
U_2(\lambda)&=U_{\mathrm{bond}}+U_{\mathrm{angle}}+U_{\mathrm{LJ}}+\lambda U_{\mathrm{target}},\\
U_3(\lambda)&=U_{\mathrm{target}}+(1-\lambda)(U_{\mathrm{bond}}+U_{\mathrm{angle}}+U_{\mathrm{LJ}}).
\end{align}
Here $U_{\mathrm{bond}}$ is the O--H spring potential in the molecular reference state. The angular restraint is
\begin{equation}
U_{\mathrm{angle}}
=
\sum_{\alpha}
k_{\theta}
\left(
\theta_{\alpha}
-
\theta_0
\right)^2,
\end{equation}
where $\theta_{\alpha}$ is the H-O-H angle of molecule $\alpha$, $\theta_0$ is the reference angle, and $k_{\theta}$ is the angular spring constant. These intermediate states are introduced to keep the path reversible and to avoid sharply varying HTI integrands.


\subsection{Consistency Checks}
Because free energy is a state function, the final integrated free energy should be independent of the thermodynamic integration path when the path is reversible and the ensemble averages are sufficiently converged. This path independence provides a useful consistency check for practical phase diagram calculations. In practice, one can perform HTI at two different temperatures under the same pressure and then use TTI to propagate the two results, yielding two independent $G(T)$ curves. Equivalently, one can perform HTI at two different pressures under the same temperature and then use pTI to obtain two independent $G(p)$ curves. Agreement between the resulting curves indicates that the HTI, TTI, or pTI calculations are mutually consistent, whereas significant discrepancies signal insufficient sampling, an irreversible integration path, or an inaccurate numerical integration.

\subsection{Numerical Integration and Error Estimation}
At the end of an HTI, TTI, or pTI calculation, \texttt{dpti} provides a post-processing module called \texttt{compute} to average the physical quantities obtained from the MD tasks and perform the numerical integration. Two quadrature schemes are supported: trapezoidal and Simpson integration. The default is Simpson integration, and the quadrature scheme can be selected from the command line using the \texttt{compute --scheme} option, for example \texttt{--scheme simpson} or \texttt{--scheme trapezoidal}.

Together with the integrated free energy, \texttt{dpti} reports two uncertainties: a statistical error and an integration error. The statistical error is reported together with the integrated free energy to provide the error bar of the free energy estimate. The integration error is an indicator of the numerical quadrature error and is used to guide the refinement of the integration grid.

The statistical error comes from block averaging of the MD trajectory at each integration point and is propagated through the numerical quadrature. If the thermodynamic integral is written as
\begin{equation}
I = \sum_i w_i f_i,
\end{equation}
where \(f_i\) is the averaged integrand at the \(i\)-th integration point and \(w_i\) is the quadrature weight associated with the selected integration scheme, then, assuming that the statistical errors at different integration points are independent, the statistical uncertainty is estimated as
\begin{equation}
\sigma_I =
\left[
\sum_i
\left(
w_i \sigma_i
\right)^2
\right]^{1/2},
\end{equation}
where \(\sigma_i\) is the block-averaged statistical uncertainty of \(f_i\). Statistical errors from independent integration segments are combined in quadrature.

The integration error estimates the discretization error of the numerical quadrature. For each interval \([\lambda_i,\lambda_{i+1}]\), \texttt{dpti} estimates the local curvature of the integrand, \(f''\), from nearby integration points. For an interior interval, two curvatures are evaluated by fitting quadratic interpolants to the adjacent three-point stencils: \(f''_{i-1,i,i+1}\) from \(\lambda_{i-1}\), \(\lambda_i\), and \(\lambda_{i+1}\), and \(f''_{i,i+1,i+2}\) from \(\lambda_i\), \(\lambda_{i+1}\), and \(\lambda_{i+2}\). The larger absolute curvature is used for the interval,
$
\left|f''\right|
=
\max
\left(
\left|f''_{i-1,i,i+1}\right|,
\left|f''_{i,i+1,i+2}\right|
\right)$,
and the interval error is estimated as
\begin{equation}\label{e_i_i+1}
\epsilon_{i,i+1}
\approx
\frac{
\left(\lambda_{i+1}-\lambda_i\right)^3
\left|f''\right|
}{12}.
\end{equation}
At the two ends of the integration path, the one-sided three-point stencil is used. The estimated interval errors are summed along the integration path to obtain the reported integration error.

The estimated $\epsilon_{i, i+1}$ helps to determine whether more integration points are needed in the interval $[\lambda_i, \lambda_{i+1}]$. Given a target total error $\epsilon^{\text{tot}}$, \texttt{dpti} assigns each interval an error tolerance $\epsilon_{i, i+1}^{tol}=\epsilon^{\text{tot}}(\lambda_{i+1}-\lambda_i)$ proportional to its length and divides the interval into \(n_i\) subintervals according to
\begin{equation}\label{n_i}
n_i =
\max\left[
1,
\left\lceil
\sqrt{\frac{\epsilon_{i, i+1}}{\epsilon_{i, i+1}^{tol}}}
\right\rceil
\right].
\end{equation}
\texttt{dpti} adds \(n_i-1\) new grid points in subintervals $[\lambda_i, \lambda_{i+1}]$ where the estimated error exceeds the assigned tolerance.

Finite-size errors~\cite{polson_finite-size_2000} provide another source of uncertainty that is not estimated automatically by \texttt{dpti}. In a CALPHY calculation for bcc Fe, systems with on the order of \(10^3\) atoms gave finite-size errors of about 1~meV/atom, while systems with more than \(10^4\) atoms reduced the error below 0.1~meV/atom~\cite{menon_automated_2021}. Because MLIP simulations can commonly use cells containing several thousand atoms, this error is often small for phase-boundary calculations.

\section{Software Usage}
\texttt{dpti} is available on GitHub~\cite{noauthor_httpsgithubcomdeepmodelingdpti_2024} under the LGPL-3.0 license and can be installed using \texttt{pip}.
\texttt{dpti} requires users to prepare an initial configuration for each phase of interest and \texttt{json} files specifying the simulation conditions for each stage of the workflow. Here, ``stage'' refers to an operation within the overall \texttt{dpti} workflow, whereas ``step'' denotes a sub-part within an HTI path. For the first four stages, a \texttt{json} file is needed for each phase of each stage, such as \texttt{npt.ice.json}, \texttt{hti.water.json}, etc. For the fifth stage, GDI, two \texttt{json} files are needed for each pair of phases: one specifying MD simulation parameters and the other specifying the GDI path. Users also need to prepare an MLIP model. Moreover, \texttt{machine.json} files can be provided to take advantage of \texttt{DPDispatcher} for managing the execution of MD tasks on HPC clusters. The meanings of commonly used JSON parameters are summarized in Table~S1 of the Supporting Information.

We demonstrate the workflow for phase diagram calculations using \texttt{dpti} with two pedagogical examples, silica and water. We first consider silica as a standard atomic system and assemble several pairwise phase boundaries into a phase diagram. We then consider water as a molecular system that requires a specialized reference state and HTI path. The extension of this workflow to other phases is straightforward. All input files can be found in the accompanying GitHub repository~\cite{li_httpsgithubcomyi-fanlidpti_manuscript_examples_nodate}.
\subsection{Example: Silica}
This example showcases how to calculate the phase boundaries between $\beta$-quartz, coesite, and melt SiO$_2$, as in the phase diagram reported in Ref.~\citenum{zhong_general_2026}. We use the DP model from Ref.~\citenum{zhong_general_2026}, which was trained on DFT data computed with the R2SCAN exchange-correlation functional~\cite{furness_accurate_2020} and is referred to here as DP@R2SCAN. All MD tasks in this example use a timestep of 2~fs. Liquid phases require longer runs because of slower structural relaxation.

\subsubsection{Stage 1: $NpT$ Simulation}
The first stage determines the equilibrium cell of each phase at the chosen state point $(p_0,T_0)$. \texttt{dpti} takes a LAMMPS data file, \texttt{conf.lmp}, as the initial configuration and an input file, \texttt{npt.json}, to specify simulation conditions including the ensemble, timestep, and simulation length. For coesite and $\beta$-quartz, we use the anisotropic $NpT$ ensemble by setting \texttt{ens} to \texttt{npt-aniso}; for the melt, we use the isotropic $NpT$ ensemble by setting \texttt{ens} to \texttt{npt}. The LAMMPS input file is generated with
\commandboxneedspace
\begin{center}
\begin{minipage}{0.5\textwidth}
\begin{grayverb}
\begin{spverbatim}
dpti equi gen npt.json -o npt
\end{spverbatim}
\end{grayverb}
\end{minipage}
\end{center}
\vspace{1em}

\texttt{dpti equi run} then submits the MD task to HPC clusters through \texttt{DPDispatcher}~\cite{yuan_dpdispatcher_2025}, the task-dispatching backend also used by DP-GEN~\cite{zhang_dp-gen_2020}. The \texttt{machine.gpu.json} file specifies the HPC settings and should be adapted to the user's machine:
\commandboxneedspace
\begin{center}
\begin{minipage}{0.5\textwidth}
\begin{grayverb}
\begin{spverbatim}
dpti equi run npt machine.gpu.json
\end{spverbatim}
\end{grayverb}
\end{minipage}
\end{center}
\vspace{1em}

After the simulation finishes, \texttt{dpti equi compute} extracts the averaged thermodynamic quantities:
\commandboxneedspace
\begin{center}
\begin{minipage}{0.5\textwidth}
\begin{grayverb}
\begin{spverbatim}
dpti equi compute npt
\end{spverbatim}
\end{grayverb}
\end{minipage}
\end{center}
\vspace{1em}

The resulting equilibrium density and cell parameters are used in the subsequent $NVT$ and HTI calculations. In this example, the $\beta$-quartz and coesite calculations use $5\times5\times5$ and $3\times2\times3$ supercells, corresponding to 1125 and 864 atoms, respectively, while the melt calculations use 648 atoms. The solid $NpT$ simulations are run for 200~ps, whereas the melt simulations are run for 4~ns to obtain stable liquid densities. The state points and averaged equilibrium quantities are summarized in Table~\ref{tab:silica_npt_results}.

\begin{table}[htbp]
\centering
\caption{\(NpT\) simulation results for the silica example}
\label{tab:silica_npt_results}
\begin{tabular}{lcccccccc}
\toprule
Phase & 
\(T_0\) [K] & 
\(p_0\) [bar] & 
\(\rho\) [g/cm\(^3\)] & 
\(L_x\) [\AA] & 
\(L_y\) [\AA] & 
\(L_z\) [\AA] & 
\(L_{xz}\) [\AA] & 
\(L_{xy}\) [\AA] \\
\midrule
Coesite            & 1600 & 10000 & 2.838 & 21.75 & 24.95 & 18.66 & -10.92 & 0.00 \\
Coesite            & 1600 & 50000 & 2.932 & 21.45 & 24.70 & 18.49 & -10.82 & 0.00 \\
Coesite            & 2000 & 50000 & 2.921 & 21.49 & 24.74 & 18.51 & -10.83 & 0.00 \\
\(\beta\)-Quartz   & 1600 & 10000 & 2.536 & 25.02 & 21.67 & 27.21 & 0.00 & -12.51 \\
\(\beta\)-Quartz   & 1600 & 50000 & 2.751 & 24.28 & 21.03 & 26.64 & 0.00 & -12.14 \\
\(\beta\)-Quartz   & 2000 & 50000 & 2.671 & 24.58 & 21.29 & 26.76 & 0.00 & -12.29 \\
Melt               & 3200 & 50000 & 2.629 & 20.16 & 20.16 & 20.16 & 0.00 & 0.00 \\
Melt               & 3300 & 50000 & 2.626 & 20.17 & 20.17 & 20.17 & 0.00 & 0.00 \\
\bottomrule
\end{tabular}
\end{table}

\subsubsection{Stage 2: $NVT$ Simulation}
The $NVT$ stage equilibrates each phase in the fixed cell obtained from the preceding $NpT$ simulation to generate the initial configuration for the subsequent HTI calculation. The option \texttt{--conf-npt npt} tells \texttt{dpti} to take both the equilibrated box and the final configuration from the Stage~1 \texttt{npt} directory as the initial box and configuration for the $NVT$ task. For crystalline phases, we recommend setting \texttt{if\_dump\_avg\_posi} to \texttt{true} in \texttt{nvt.json}, so that trajectory-averaged atomic positions are written out and used as more stable reference positions for HTI. For liquid phases, this option must be set to \texttt{false}, because averaging diffusive atomic coordinates would not produce a meaningful liquid configuration. In this example, each $NVT$ equilibration is run for 100~ps. The following commands generate and run the $NVT$ tasks:
\commandboxneedspace
\begin{center}
\begin{minipage}{0.5\textwidth}
\begin{grayverb}
\begin{spverbatim}
dpti equi gen nvt.json -o nvt \
  --conf-npt npt
dpti equi run nvt machine.gpu.json
\end{spverbatim}
\end{grayverb}
\end{minipage}
\end{center}
\vspace{1em}

\subsubsection{Stage 3: Hamiltonian Thermodynamic Integration}

HTI connects the reference and target states by switching $\lambda$ from 0 to 1. The $\lambda$ grid is specified in the \texttt{json} file and can be coarse in Stage 3 because \texttt{dpti} adaptively refines it in Stage 4 to reduce the integration error. Thus, users do not need to fine-tune the initial $\lambda$ grid.

\paragraph{Coesite and $\beta$-quartz}
For the crystalline phases, coesite and $\beta$-quartz, one-step HTI, which turns on the DP potential and switches off the spring potential simultaneously, is performed with the generic solid-state \texttt{dpti hti} workflow. For each crystalline phase, we perform HTI at $T_0=1600$ and 2000~K under $p_0=5$~GPa to provide the two anchor free energies used in the TTI consistency check. We also perform HTI at $p_0=1$ and 5~GPa under $T_0=1600$~K to provide the two anchor free energies used in the pTI consistency check. We set \texttt{langevin} to \texttt{true} because a Nos\'e--Hoover chain (NHC) thermostat can have ergodicity issues for nearly harmonic solid-state reference systems. Each MD task in HTI spans 200~ps.

In this example, we use \texttt{spring\_k}=0.15 for the Frenkel-type Einstein crystal. The MSD of coesite at 5 GPa and 1600~K from the 200-ps MD simulation is about \(0.113~\mbox{\AA}^2\). Using this all-atom MSD as a rough estimate for the species-resolved MSDs gives \(k_{\mathrm{Si}}\approx k_{\mathrm{O}}\approx 3k_{\mathrm B}T/\langle|\Delta\mathbf r|^2\rangle=3.65~\mathrm{eV}~\mbox{\AA}^{-2}\). Substituting these values into the mass-normalized estimate in the Theory section gives
\[
\kappa \approx
\left(
\frac{1}{3}\frac{m_{\mathrm{Si}}}{k_{\mathrm{Si}}}
+\frac{2}{3}\frac{m_{\mathrm{O}}}{k_{\mathrm{O}}}
\right)^{-1}
=0.18~\mathrm{eV}~\mbox{\AA}^{-2}~\mathrm{amu}^{-1},
\]
close to the value used here. Relative to \texttt{spring\_k}=0.15, the Gibbs free energy changes by less than 0.4~meV/atom over the tested range of spring constants, as shown in Fig.~S1 of the Supporting Information.

For each crystalline phase, the complete set of commands is
\commandboxneedspace
\begin{center}
\begin{minipage}{0.5\textwidth}
\begin{grayverb}
\begin{spverbatim}
dpti hti gen hti.json -o hti \
  -s one-step
dpti hti run hti \
  machine.gpu.json one-step
cd hti
dpti hti compute . \
  -t gibbs --npt ../npt
\end{spverbatim}
\end{grayverb}
\end{minipage}
\end{center}

\begin{figure}
\includegraphics[width=\textwidth]{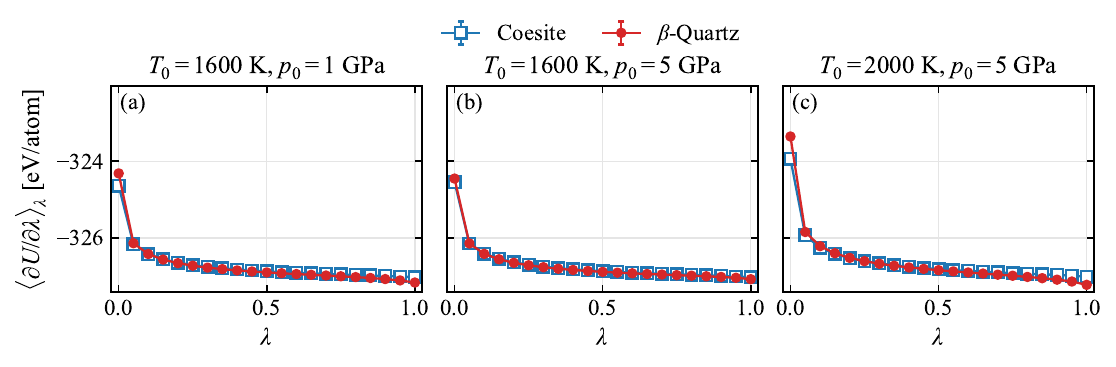}
\caption{One-step HTI integrands $\langle\partial U/\partial \lambda \rangle_{\lambda} = \langle U_{\mathrm{DP}}-U_{\mathrm{spring}} \rangle_{\lambda}$ for the crystalline silica phases, coesite and $\beta$-quartz, at the three anchor thermodynamic conditions used for the TTI and pTI consistency checks.}
\label{fig:silica_solids_hti_integrand}
\end{figure}

Figure~\ref{fig:silica_solids_hti_integrand} shows the one-step HTI integrands, $\langle \partial U/\partial \lambda \rangle_{\lambda} = \langle U_{\mathrm{DP}}-U_{\mathrm{spring}} \rangle_{\lambda}$, for coesite and $\beta$-quartz at the three anchor conditions.

The individual contributions to the final Gibbs free energy, as reported by the \texttt{dpti hti compute} post-processing command, are summarized in Table~\ref{tab:silica_solid_hti_contributions}. The final results are the Gibbs free energies per atom, $G$.

\begin{table}
\centering
\caption{Contributions to the Gibbs free energy for the crystalline silica HTI calculations shown in Fig.~\ref{fig:silica_solids_hti_integrand}. $A_0$ is the Helmholtz free energy of the harmonic reference system, $\Delta A=\int_0^1\langle\partial H(\lambda)/\partial\lambda\rangle_{\lambda}\,\mathrm{d}\lambda$ is the HTI contribution, and $\langle V\rangle$ is the average volume obtained from the preceding $NpT$ simulation. The Gibbs free energy is $G=A_0+\Delta A+p_0\langle V\rangle$. The last four columns are in $\mathrm{eV}$/atom. Numbers in parentheses denote the statistical uncertainty in the last digits.}
\label{tab:silica_solid_hti_contributions}
\small
\setlength{\tabcolsep}{3pt}
\begin{tabular}{lcccccc}
\toprule
Phase & $T_0$ [K] & $p_0$ [bar] & $A_0$ & $\Delta A$ & $p_0\langle V\rangle$ & $G$ \\
\midrule
Coesite & 1600 & 10000 & $-0.7079$ & $-326.7618(1)$ & $0.07315(1)$ & $-327.3965(1)$ \\
Coesite & 1600 & 50000 & $-0.7079$ & $-326.7543(1)$ & $0.35395(4)$ & $-327.1082(1)$ \\
Coesite & 2000 & 50000 & $-1.0002$ & $-326.6744(1)$ & $0.35537(3)$ & $-327.3192(1)$ \\
$\beta$-Quartz & 1600 & 10000 & $-0.7074$ & $-326.7905(1)$ & $0.08184(2)$ & $-327.4161(1)$ \\
$\beta$-Quartz & 1600 & 50000 & $-0.7074$ & $-326.7705(1)$ & $0.3773(1)$ & $-327.1007(2)$ \\
$\beta$-Quartz & 2000 & 50000 & $-0.9996$ & $-326.7014(1)$ & $0.3886(2)$ & $-327.3124(3)$ \\
\bottomrule
\end{tabular}
\end{table}

\paragraph{Melt}
For the liquid phase, HTI is performed with the liquid workflow \texttt{dpti hti\_liq}. This three-step HTI connects the ideal-gas reference to the target liquid through auxiliary soft-core LJ interactions, as described in the Theory section. 

\begin{figure}[htbp]
\centering
\includegraphics[width=\textwidth]{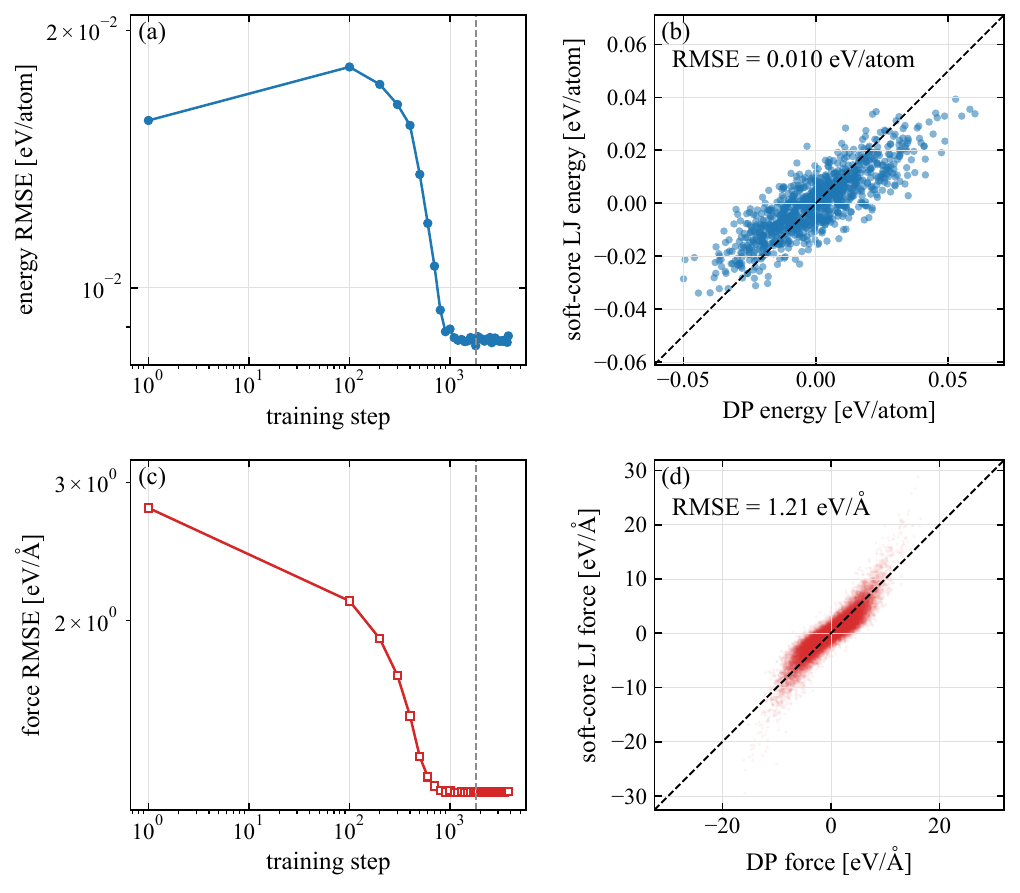}
\caption{Fitting of the soft-core LJ reference potential for silica melt. Panels (a) and (c) show the full-dataset energy and force RMSE values during optimization, respectively. Panels (b) and (d) compare the fitted soft-core LJ energies and force components against the target DP data on the 1000-frame fitting dataset.}
\label{fig:silica_melt_lj_training}
\end{figure}

Because the auxiliary LJ potential should provide a smooth and physically reasonable intermediate state, we fit its energy and length parameters to 1000 frames extracted from the final part of a 3000~K, 5~GPa silica-melt $NpT$ trajectory generated with the DP@R2SCAN potential. The fitting minimizes a weighted objective containing both centered energies and forces, and the weights are gradually shifted during training from force-dominated fitting to energy-dominated fitting: the force weight decreases from 1000 to 1, while the energy weight increases from 0.02 to 1000. Figure~\ref{fig:silica_melt_lj_training} shows the resulting full-dataset RMSE learning curves and parity plots. Although the soft-core LJ potential has limited representability, the fitted parameters capture the features of the DP potential and provide a reasonably good intermediate state for the HTI path. The fitted parameters are listed in Table~\ref{tab:silica_melt_lj_parameters}; the remaining soft-core LJ settings are $\eta=0.5$ for all pairs, $n=1$, $\alpha=0.5$, and $r_{\mathrm{cut}}=6$~\AA. The fitted parameters are consistent with chemical intuition: although long-range order is lost in silica melt, the local chemical environment remains dominated by Si--O tetrahedra, making the Si--O interaction much stronger than the Si--Si and O--O interactions.

\begin{table}
\centering
\caption{Soft-core LJ parameters used for the silica-melt HTI calculations.}
\label{tab:silica_melt_lj_parameters}
\begin{tabular}{lcc}
\toprule
Pair & $\epsilon$ [eV] & $\sigma$ [\AA] \\
\midrule
Si--Si & 0.061 & 3.385 \\
Si--O & 3.205 & 1.402 \\
O--O & 0.067 & 2.913 \\
\bottomrule
\end{tabular}
\end{table}

For the melt calculations, we set \texttt{langevin} to \texttt{false} in \texttt{hti.melt.json}, so that the simulations use the NHC thermostat. Each MD task in HTI spans 4~ns. The complete set of commands is
\commandboxneedspace
\begin{center}
\begin{minipage}{0.5\textwidth}
\begin{grayverb}
\begin{spverbatim}
dpti hti_liq gen hti.melt.json \
  -o hti -s three-step
dpti hti_liq run hti \
  machine.cpu.json 00 --no-dp
dpti hti_liq run hti \
  machine.gpu.json 01
dpti hti_liq run hti \
  machine.gpu.json 02
cd hti
dpti hti_liq compute . \
  -t gibbs --npt ../npt
\end{spverbatim}
\end{grayverb}
\end{minipage}
\end{center}

\begin{figure}[htbp]
\centering
\includegraphics[width=\textwidth]{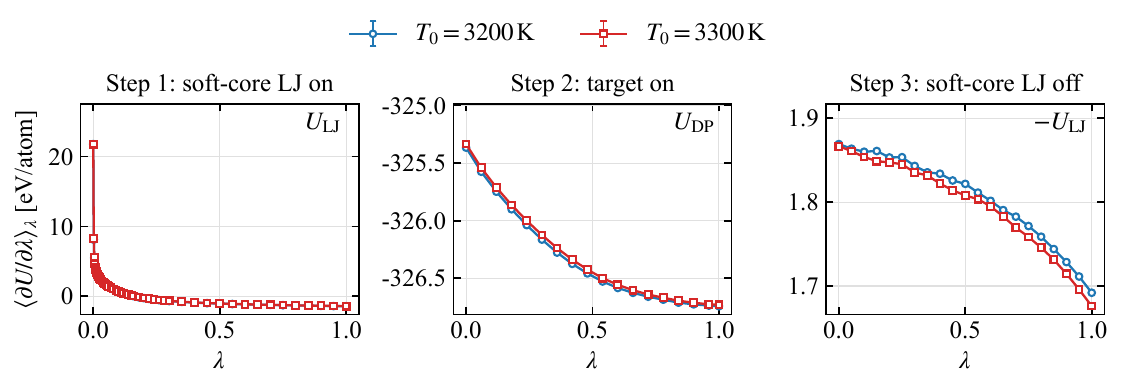}
\caption{HTI integrands $\langle\partial U/\partial\lambda\rangle_{\lambda}$ for silica melt at 3200 and 3300~K along the three-step path from the ideal-gas reference state to the target liquid. The three panels correspond to switching on the soft-core LJ interaction, switching on the target potential, and removing the auxiliary soft-core LJ interaction, respectively.}
\label{fig:silica_melt_hti_integrand}
\end{figure}

Figure~\ref{fig:silica_melt_hti_integrand} shows the three HTI integrands for the silica melt at the two liquid anchor temperatures.

The individual contributions to the final Gibbs free energy of the melt, as reported by the \texttt{dpti hti\_liq compute} post-processing command, are summarized in Table~\ref{tab:silica_melt_hti_contributions}. The results are reported per atom for the initial HTI calculations at the two liquid anchor temperatures used in the TTI consistency check.

\begin{table*}[t]
\centering
\caption{Contributions to the Gibbs free energy for the silica-melt HTI calculations. $A_0$ is the Helmholtz free energy of the ideal-gas reference system. $\Delta A_1$ is the HTI contribution from step 1, switching on the soft-core LJ interaction. $\Delta A_2$ is the HTI contribution from switching on DP and $\Delta A_3$ the contribution from switching off the soft-core LJ. $\langle V\rangle$ is the average volume obtained from the preceding $NpT$ simulation. The Gibbs free energy is $G=A_0+\Delta A_1+\Delta A_2+\Delta A_3+p_0\langle V\rangle$. The last six columns are in $\mathrm{eV}$/atom. Numbers in parentheses denote the statistical uncertainty in the last digits.}
\label{tab:silica_melt_hti_contributions}
\small
\setlength{\tabcolsep}{3pt}
\begin{tabular}{lcccccccc}
\toprule
Phase & $T_0$ [K] & $p_0$ [bar] & $A_0$ & $\Delta A_1$ & $\Delta A_2$ & $\Delta A_3$ & $p_0\langle V\rangle$ & $G$ \\
\midrule
Melt & 3200 & 50000 & $-3.3448$ & $-0.62907(3)$ & $-326.33715(6)$ & $1.80672(3)$ & $0.3948(1)$ & $-328.1095(2)$ \\
Melt & 3300 & 50000 & $-3.4627$ & $-0.60474(3)$ & $-326.30899(6)$ & $1.79699(3)$ & $0.3952(1)$ & $-328.1842(2)$ \\
\bottomrule
\end{tabular}
\end{table*}

\paragraph{Refinement of the HTI Grid}
After an initial HTI calculation, the discretization error of the $\lambda$ integration can be reduced by refining the HTI grid. The refinement procedure is necessary when the original HTI calculation uses a coarse $\lambda$ grid, which may lead to an error as large as several meV/atom in the final free energy.

The \texttt{refine} command reads the existing \texttt{hti.out} file, estimates which $\lambda$ intervals require additional points using the integration-error estimate described in the Theory section, and generates a new HTI task directory while reusing the completed tasks from the original calculation. Using $n_i$ defined in Eq.~\eqref{n_i}, \texttt{dpti} adds \(n_i-1\) new grid points in the subinterval $[\lambda_i, \lambda_{i+1}]$. The following commands generate a refined HTI task for the solid phases with a target integration error of $10^{-3}$~eV/atom, run the new tasks, and compute the refined result:
\commandboxneedspace
\begin{center}
\begin{minipage}{0.5\textwidth}
\begin{grayverb}
\begin{spverbatim}
dpti hti refine -i hti \
  -o hti.refine -e 1e-3
dpti hti run hti.refine \
  machine.gpu.json one-step
cd hti.refine
dpti hti compute . \
  -t gibbs --npt ../npt
\end{spverbatim}
\end{grayverb}
\end{minipage}
\end{center}
The \texttt{refine} command writes a \texttt{refine.out} file to record the number of new grid points added in each interval. \texttt{dpti hti run} detects which tasks have already been completed in the original HTI calculation and only runs the new tasks corresponding to the added grid points. In this example, each newly added MD task in the refined HTI grid is also run for the same length as the original tasks, i.e., 200~ps for the crystalline phases and 4~ns for the melt. After the new tasks finish, \texttt{dpti hti compute} is run again to obtain the refined free energy. 

\begin{figure}[htbp]
\centering
\includegraphics[width=\textwidth]{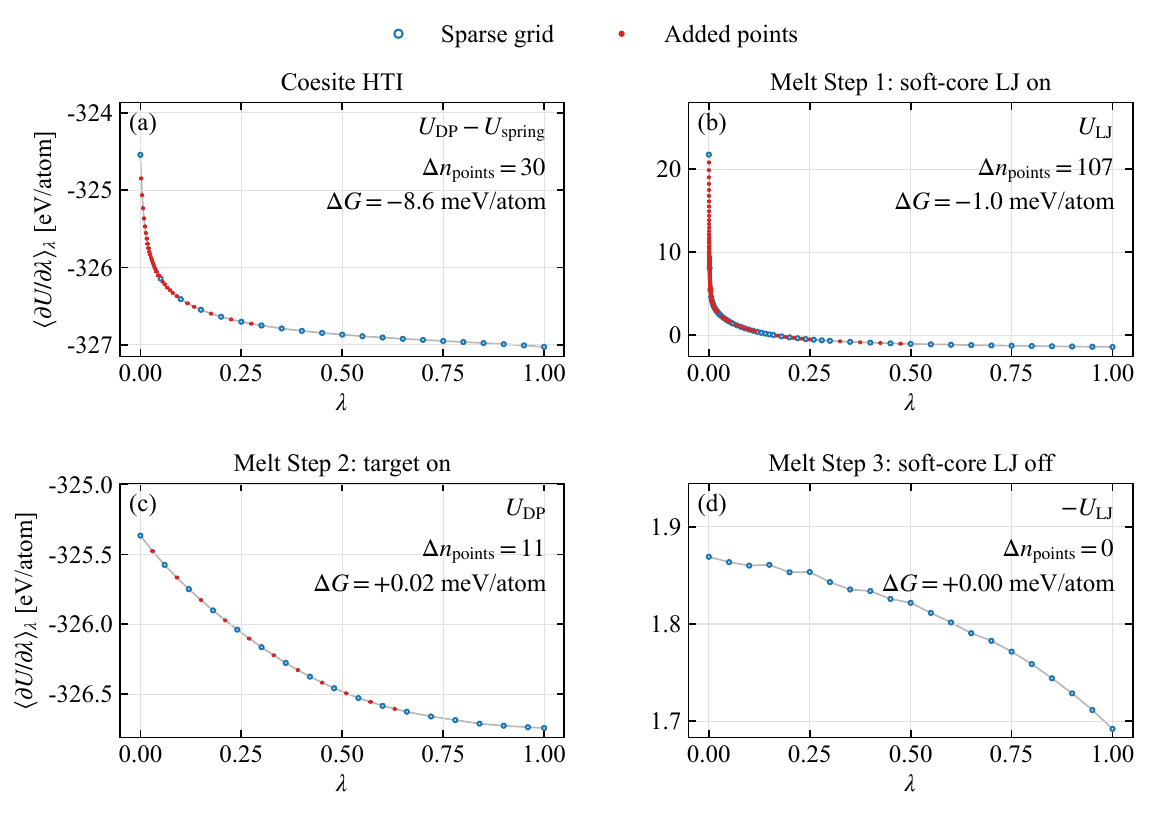}
\caption{Representative HTI-grid refinement for silica using a target total error of 1~meV/atom. Sparse points are the original $\lambda$ grids used in the initial HTI calculations, and added points are the new $\lambda$ values introduced by \texttt{dpti refine}. Panel (a) shows the one-step coesite HTI integrand at $T_0=1600$~K and $p_0=50000$~bar, and panels (b)--(d) show the three HTI steps for the silica melt at $T_0=3200$~K and $p_0=50000$~bar. The annotated $\Delta G$ values are the changes in the corresponding integrated free-energy contributions after refinement.}
\label{fig:silica_hti_refinement_sparse_added}
\end{figure}

A similar set of commands can be used to refine the HTI grid for the melt phase:
\commandboxneedspace
\begin{center}
\begin{minipage}{0.5\textwidth}
\begin{grayverb}
\begin{spverbatim}
dpti hti_liq refine -i hti \
  -o hti.refine -e 1e-3
dpti hti_liq run hti.refine \
  machine.cpu.json 00 --no-dp
dpti hti_liq run hti.refine \
  machine.gpu.json 01
dpti hti_liq run hti.refine \
  machine.gpu.json 02
cd hti.refine
dpti hti_liq compute . \
  -t gibbs --npt ../npt
\end{spverbatim}
\end{grayverb}
\end{minipage}
\end{center}
Figure~\ref{fig:silica_hti_refinement_sparse_added} shows representative examples of the sparse grid points and the additional points introduced by refinement.

Table~\ref{tab:silica_hti_refinement_comparison} compares the Gibbs free energies before and after HTI-grid refinement for the silica example. The refinement changes the melt free energies by about 1~meV/atom, while the corrections for the crystalline phases are larger for the coarse initial grids used here.

\begin{table*}[t]
\centering
\caption{Effect of HTI-grid refinement on the Gibbs free energy. $G_{\mathrm{before}}$ and $G_{\mathrm{after}}$ are in eV/atom. $\Delta n_{\mathrm{points}}$ is the number of new $\lambda$-grid points introduced by refinement, summed over all HTI steps for the melt. $\Delta G=G_{\mathrm{after}}-G_{\mathrm{before}}$ is the difference between the Gibbs free energies after and before refinement.}
\label{tab:silica_hti_refinement_comparison}
\begin{tabular}{lcccccc}
\toprule
Phase & $T_0$ [K] & $p_0$ [bar] & $G_{\mathrm{before}}$ & $G_{\mathrm{after}}$ & $\Delta n_{\mathrm{points}}$ & $\Delta G$ [meV/atom] \\
\midrule
Coesite & 1600 & 10000 & $-327.3965(1)$ & $-327.4048(1)$ & 26 & $-8.2$ \\
Coesite & 1600 & 50000 & $-327.1082(1)$ & $-327.1168(1)$ & 27 & $-8.6$ \\
Coesite & 2000 & 50000 & $-327.3192(1)$ & $-327.3299(1)$ & 30 & $-10.7$ \\
$\beta$-Quartz & 1600 & 10000 & $-327.4161(1)$ & $-327.4264(1)$ & 31 & $-10.3$ \\
$\beta$-Quartz & 1600 & 50000 & $-327.1007(2)$ & $-327.1099(1)$ & 29 & $-9.1$ \\
$\beta$-Quartz & 2000 & 50000 & $-327.3124(3)$ & $-327.3259(2)$ & 36 & $-13.4$ \\
Melt & 3200 & 50000 & $-328.1095(2)$ & $-328.1104(1)$ & 118 & $-0.9$ \\
Melt & 3300 & 50000 & $-328.1842(2)$ & $-328.1853(1)$ & 117 & $-1.0$ \\
\bottomrule
\end{tabular}
\end{table*}

\subsubsection{Stage 4: Temperature and Pressure Thermodynamic Integration}
For silica, both temperature thermodynamic integration (TTI) and pressure thermodynamic integration (pTI) are used in Stage~4. For the two crystalline phases, coesite and $\beta$-quartz, we perform both TTI and pTI so that their Gibbs free energies can be propagated along both isobars and isotherms. The coesite--$\beta$-quartz phase boundary is then located from pTI because the corresponding coexistence line has a small $\partial p/\partial T$ and is therefore more conveniently determined from free-energy crossings along isotherms. For the melt, only TTI is performed. The solid--liquid phase boundary is then located from TTI, i.e., from the crossing between the Gibbs free energies of the crystalline phase and the melt along an isobar.

For a representative TTI calculation, the commands are
\commandboxneedspace
\begin{center}
\begin{minipage}{0.5\textwidth}
\begin{grayverb}
\begin{spverbatim}
dpti ti gen ti.t.json -o ti.t
dpti ti run ti.t machine.gpu.json
dpti ti compute . -H ../hti
\end{spverbatim}
\end{grayverb}
\end{minipage}
\end{center}
\vspace{1em}

Here, \texttt{-H} specifies the HTI task directory that provides the anchor Gibbs free energy at the starting state point. The pTI calculations and the TTI calculations for other phases use analogous commands, with \texttt{ti.p.json} or the corresponding phase-specific \texttt{ti.t.json} input file and HTI directory.

\begin{figure}[htbp]
\centering
\includegraphics[width=\textwidth]{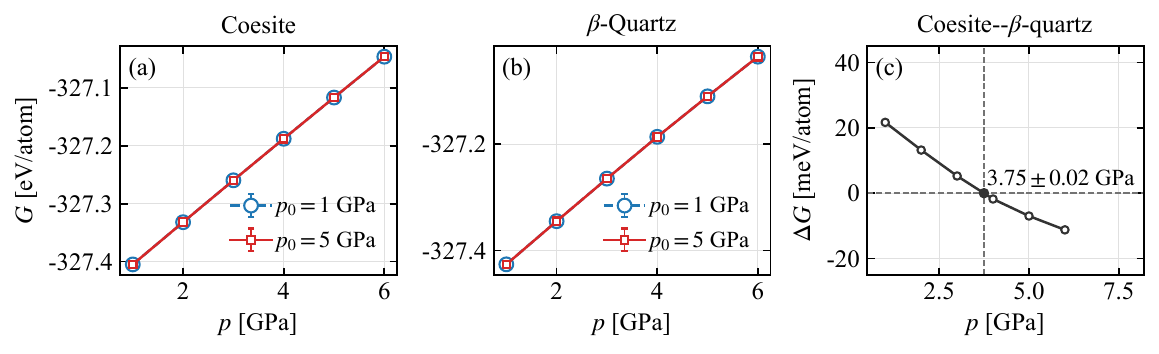}
\caption{Pressure thermodynamic integration for crystalline silica at $T=1600$~K. (a,b) Gibbs free energies of coesite and $\beta$-quartz propagated from $p_0=1$ and 5~GPa. The agreement between the two propagated curves provides a pTI consistency check. (c) Representative free-energy difference $\Delta G=G_{\mathrm{coesite}}-G_{\beta\mathrm{-quartz}}$ from the $p_0=5$~GPa curves, giving $p_{\mathrm{tr}}=3.75\pm0.02$~GPa.}
\label{fig:silica_pti_g_diff}
\end{figure}

Figure~\ref{fig:silica_pti_g_diff} shows the pTI results for coesite and $\beta$-quartz at $T=1600$~K. The solid-phase pTI and TTI simulations are run for 200~ps at each pressure or temperature point. The first two panels compare the Gibbs free energies propagated from the two anchor pressures, $p_0=1$ and 5~GPa, providing a consistency check for the pTI calculations; the corresponding differences between the two propagated curves are shown in Fig.~S2 of the Supporting Information. The right panel shows one representative free-energy difference curve, $\Delta G=G_{\mathrm{coesite}}-G_{\beta\mathrm{-quartz}}$, whose zero crossing gives the coesite--$\beta$-quartz transition pressure, which is $3.75\pm0.02$~GPa at 1600~K. $\Delta G$ has a statistical error of approximately 0.16~meV/atom near the transition point and translates into an uncertainty of 0.02~GPa in the transition pressure.

\begin{figure}[htbp]
\centering
\includegraphics[width=\textwidth]{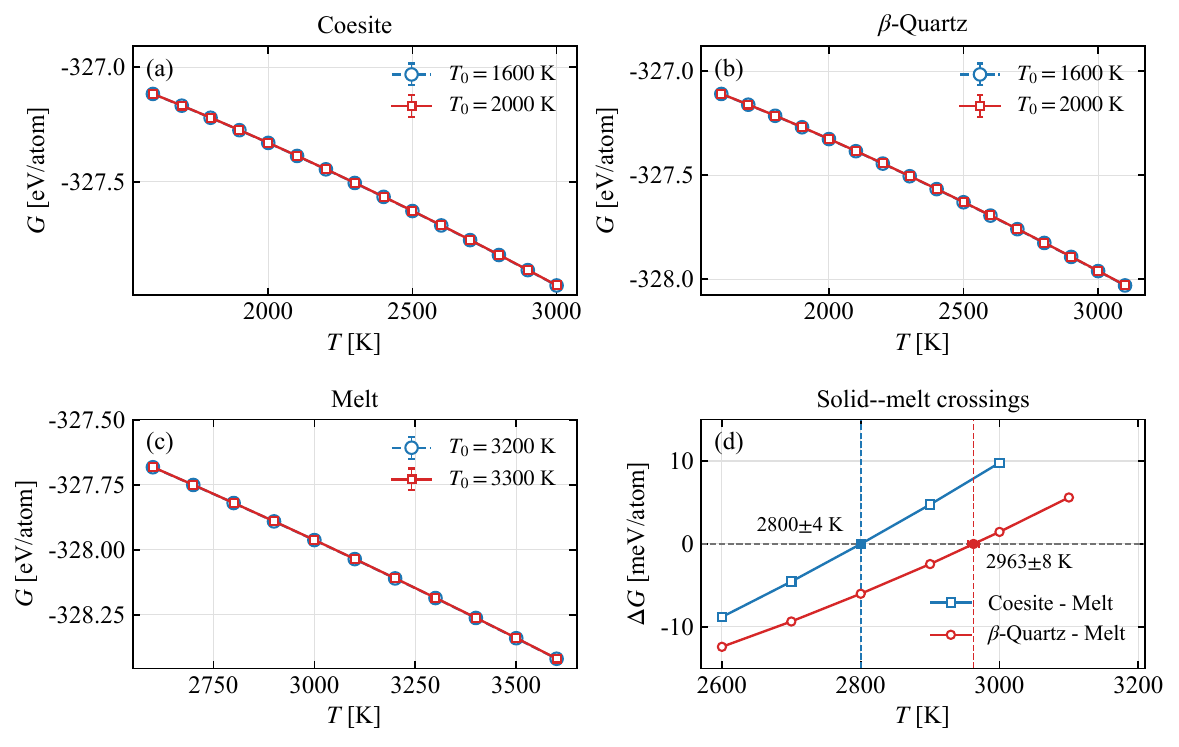}
\caption{Temperature thermodynamic integration for silica at \(p=5\)~GPa using refined HTI anchors. (a--c) Gibbs free energies propagated from two HTI temperatures for coesite, $\beta$-quartz, and the melt, respectively. (d) Representative free-energy differences for coesite--melt and $\beta$-quartz--melt. The zero crossings give transition temperatures of $T_m=2800\pm4$~K for coesite--melt and $2963\pm8$~K for $\beta$-quartz--melt.}
\label{fig:silica_tti_consistency_solid_melt}
\end{figure}

Figure~\ref{fig:silica_tti_consistency_solid_melt} shows the corresponding TTI calculations at \(p=5\)~GPa. For each phase, the free-energy curves propagated from two independently computed HTI anchors agree with each other, providing a consistency check for the TTI workflow; the corresponding anchor-to-anchor differences are shown in Figs.~S3 and S4 of the Supporting Information. For the melt, each TTI MD simulation is run for 4~ns to converge the liquid enthalpy. The representative solid--melt free-energy differences in panel (d) give melting temperatures of $2800\pm4$~K for coesite--melt and $2963\pm8$~K for $\beta$-quartz--melt. The statistical errors of the free energy differences, 0.2 meV/atom for coesite--melt and 0.3 meV/atom for $\beta$-quartz--melt, translate into uncertainties of 4~K and 8~K in the corresponding melting temperatures, respectively.

\subsubsection{Stage 5: Gibbs-Duhem Integration}
\begin{figure}[htbp]
\centering
\includegraphics[width=\textwidth]{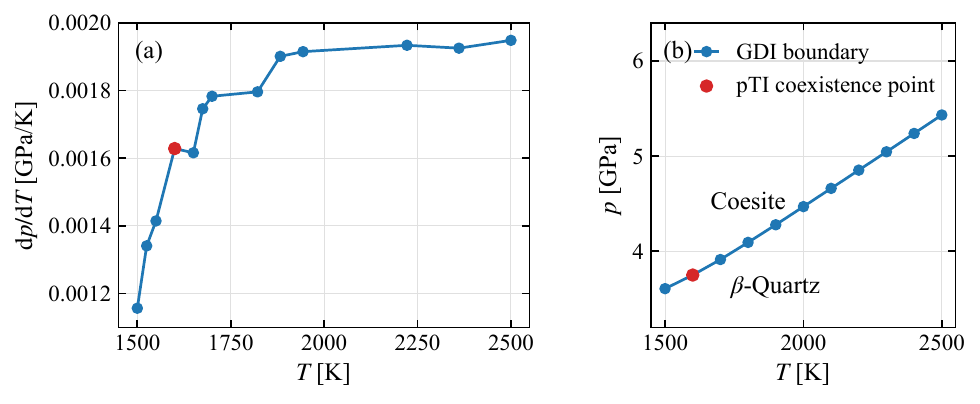}
\caption{Gibbs-Duhem integration of the coesite--$\beta$-quartz phase boundary. (a) The slope \(\mathrm{d}p/\mathrm{d}T\) evaluated from the MD simulations. (b) The propagated phase-boundary points. The red markers denote the coexistence point at \(T=1600\)~K and 3.75~GPa obtained from pTI and used to initialize the GDI calculation.}
\label{fig:silica_gdi_phase_boundary}
\end{figure}

Starting from the coesite--$\beta$-quartz coexistence point 3.75~GPa obtained by pTI at \(T=1600\)~K, GDI propagates the solid--solid phase boundary in the \(T\)-\(p\) plane. Each phase-pair MD simulation used to evaluate \(\mathrm{d}p/\mathrm{d}T\) is run for 200~ps. The evaluated \(\mathrm{d}p/\mathrm{d}T\) values are written to \texttt{dpdt.out}, and the propagated phase-boundary points are written to \texttt{pb.out}. Figure~\ref{fig:silica_gdi_phase_boundary} summarizes both outputs: panel (a) shows the evaluated \(\mathrm{d}p/\mathrm{d}T\) points, and panel (b) shows the resulting boundary from 1500 to 2500~K.

The GDI calculations are launched with \texttt{dpti gdi}. The file \texttt{in.gdi.json} defines the two phases and MD settings, while \texttt{gdidata.json} defines the initial coexistence point, propagation direction, and target points. Here, the coesite--$\beta$-quartz boundary is propagated along the \(T\) path, while the two solid--melt boundaries are propagated along the \(p\) path. The command is
\commandboxneedspace
\begin{center}
\begin{minipage}{0.5\textwidth}
\begin{grayverb}
\begin{spverbatim}
dpti gdi in.gdi.json \
  machine.gpu.json -g gdidata.json
\end{spverbatim}
\end{grayverb}
\end{minipage}
\end{center}

\begin{figure}[htbp]
\centering
\includegraphics[width=\textwidth]{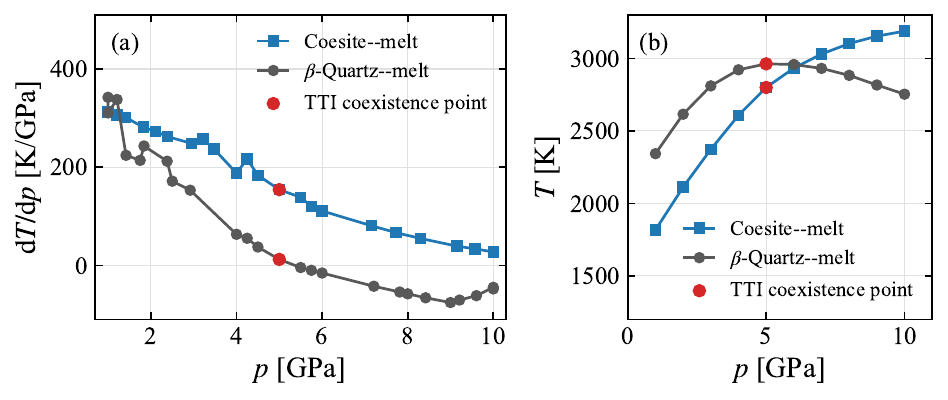}
\caption{Gibbs-Duhem integration of the silica solid--melt phase boundaries. (a) The slope \(\mathrm{d}T/\mathrm{d}p\) evaluated from the MD simulations. (b) The propagated coesite--melt and $\beta$-quartz--melt phase-boundary points. The red markers denote the coexistence points at \(p=5\)~GPa obtained from TTI and used to initialize the GDI calculations.}
\label{fig:silica_gdi_solid_melt_boundaries}
\end{figure}

Starting from the solid--melt coexistence points, 2800~K for coesite--melt and 2963~K for $\beta$-quartz--melt, obtained by TTI at \(p=5\)~GPa, GDI similarly propagates the coesite--melt and $\beta$-quartz--melt phase boundaries. The phase-pair MD simulations for these solid--melt slope evaluations are run for 2~ns. Figure~\ref{fig:silica_gdi_solid_melt_boundaries} shows the corresponding slope evaluations and propagated boundaries. These two solid--melt boundaries, together with the coesite--$\beta$-quartz boundary in Fig.~\ref{fig:silica_gdi_phase_boundary}, provide the pairwise boundaries used to assemble the local silica phase diagram.

\begin{figure}[htbp]
\centering
\begin{minipage}{0.48\textwidth}
\centering
\includegraphics[width=\textwidth]{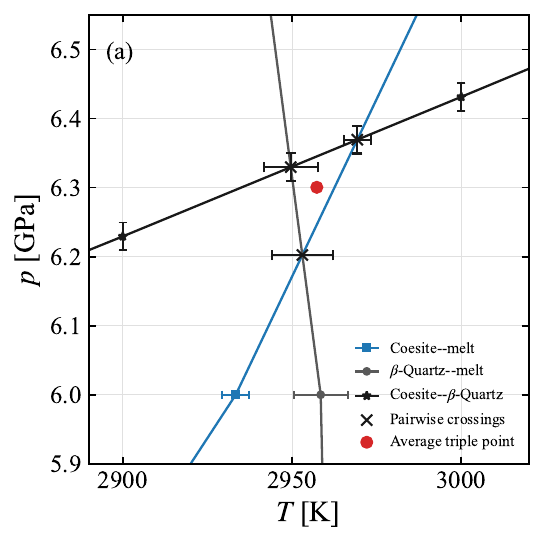}
\end{minipage}
\begin{minipage}{0.48\textwidth}
\centering
\includegraphics[width=\textwidth]{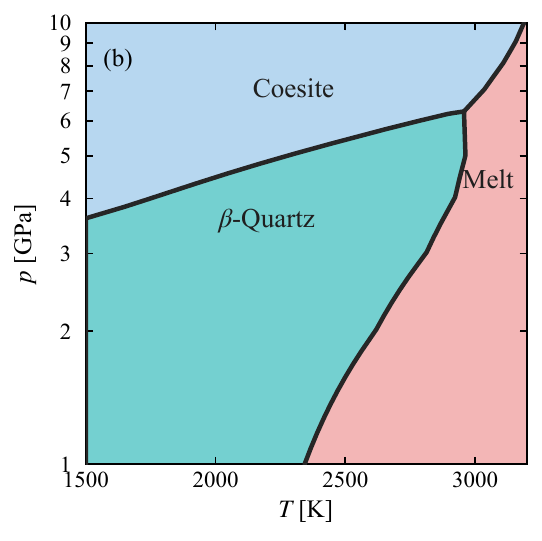}
\end{minipage}
\caption{Local silica phase diagram constructed from the propagated GDI phase-boundary points in \texttt{pb.out}. (a) Zoomed-in view of the triple-point region, showing the raw propagated pairwise boundaries without phase-field coloring. (b) Full local phase diagram in the plotted \(T\)-\(p\) window. The colored regions indicate the stable coesite, $\beta$-quartz, and melt phases.}
\label{fig:silica_phase_diagram_region}
\end{figure}

As shown in Figure~\ref{fig:silica_phase_diagram_region}(a), the coesite--melt and $\beta$-quartz--melt boundaries intersect at \(T=2953\)~K and \(p=6.20\)~GPa. The coesite--melt and coesite--$\beta$-quartz boundaries intersect at \(T=2969\)~K and \(p=6.37\)~GPa, while the $\beta$-quartz--melt and coesite--$\beta$-quartz boundaries intersect at \(T=2950\)~K and \(p=6.33\)~GPa. Averaging these three pairwise intersections gives a triple point of \(T=2957\)~K and \(p=6.30\)~GPa. This averaged triple point agrees with the three pairwise crossing estimates within the error bars shown in Fig.~\ref{fig:silica_phase_diagram_region}(a), demonstrating the mutual consistency of the independently propagated GDI boundaries. These error bars are approximated from the pTI and TTI uncertainties in Figs.~\ref{fig:silica_pti_g_diff}(c) and \ref{fig:silica_tti_consistency_solid_melt}(d); they do not include uncertainty accumulated during GDI propagation and are therefore underestimated. Combining these boundaries then yields the local \(p\)-\(T\) phase diagram shown in Fig.~\ref{fig:silica_phase_diagram_region}(b). The determined triple point is close to the previously reported triple point for $\beta$-quartz--coesite--melt from experiment at ca. 2700 K and 4.5 GPa~\cite{swamy_thermodynamic_1994}.

\subsection{Example: Water}
Here we provide a hands-on guide for calculating the phase boundary between ice Ih and liquid water. We employ the DP model trained on DFT with the revPBE-D3 functional (hereafter called DP@revPBE-D3) in Ref.~\citenum{li_ab_2025}. All MD tasks in this example use a timestep of 0.5~fs.

The first three stages described below yield the Gibbs free energy $G$ of a single phase at a given pressure and temperature $(p_0, T_0)$. In this example, we consider a fixed pressure of $p_0 = 1$~bar. For each phase, we carry out the workflow at two HTI temperatures for a consistency check: 150 and 300~K for ice Ih, and 300 and 350~K for liquid water. Stage~4 then propagates the free energy from each $T_0$ to obtain the ice Ih--liquid water coexistence temperature, and Stage~5 propagates the phase boundary by Gibbs-Duhem integration (GDI) starting from the coexistence point.
\subsubsection{Stage 1: $NpT$ Simulation}
This stage yields the equilibrium box sizes of ice Ih and liquid water at $(p_0, T_0)$ by performing an MD simulation in the $NpT$ ensemble. \texttt{dpti} requires an input file \texttt{npt.json} to specify the job and a LAMMPS data file \texttt{conf.lmp} for the initial configuration. In \texttt{npt.json}, we set \texttt{ens} to \texttt{npt-aniso} for ice and \texttt{npt} for water, corresponding to the anisotropic and isotropic $NpT$ ensembles for the solid and liquid phases, respectively. This stage is performed with the following commands:
\commandboxneedspace
\begin{center}
\begin{minipage}{0.5\textwidth}
\begin{grayverb}
\begin{spverbatim}
dpti equi gen npt.json -o npt
dpti equi run npt machine.gpu.json
dpti equi compute npt
\end{spverbatim}
\end{grayverb}
\end{minipage}
\end{center}
\vspace{1em}

We perform all $NpT$ MD simulations using a box of 432 H$_2$O molecules for 1~ns. The quantities obtained in this stage and used in subsequent calculations are summarized in Table~\ref{step1_npt_results}.

\begin{table}
\centering
\caption{$NpT$ simulation results used in the subsequent workflow}
\label{step1_npt_results}
\begin{tabular}{ccccccc}\toprule
Phase & $T_0$ [K]& $p_0$ [bar]& $\rho$ [g/cm$^3$]& $L_x$ [\AA]& $L_y$ [\AA] & $L_z$ [\AA]\\
\midrule
Ice Ih & 150 & 1 & 0.916 & 23.4608 & 22.2273 & 27.0559 \\
Ice Ih & 300 & 1 & 0.885 & 23.7248 & 22.4997 & 27.3501 \\
Water & 300 & 1 & 0.920 & 24.1291 & 24.1291 & 24.1291 \\
Water & 350 & 1 & 0.903 & 24.2784 & 24.2784 & 24.2784 \\
\bottomrule
\end{tabular}
\end{table}

\subsubsection{Stage 2: $NVT$ Simulation}
The $NVT$ stage is performed with the following commands:

\begin{center}
\begin{minipage}{0.5\textwidth}
\begin{grayverb}
\begin{spverbatim}
dpti equi gen nvt.json -o nvt \
  --conf-npt npt
dpti equi run nvt machine.gpu.json
\end{spverbatim}
\end{grayverb}
\end{minipage}
\end{center}
\vspace{1em}

For each phase, the $NVT$ equilibration was run for 100~ps to generate an equilibrated configuration for the subsequent HTI calculation.

\subsubsection{Stage 3: Hamiltonian Thermodynamic Integration}
This stage computes the absolute Gibbs free energy of the target system at $p_0=1$ bar and $T_0$ by performing HTI from a reference system. For ice, we perform HTI at $T_0=150$~K and 300~K, and for water we perform HTI at $T_0=300$~K and 350~K.

For ice, the HTI tasks are generated from \texttt{hti.ice.json}. This file specifies the settings for the HTI workflow, including the $\lambda$ grids for the three HTI steps, the spring constant \texttt{spring\_k} of the Einstein crystal, the crystal type (\texttt{frenkel} or \texttt{vega}), and the thermostat settings. In this example, we use the Frenkel formulation, although \texttt{dpti} also supports the Vega formulation. We also set \texttt{langevin} to \texttt{true}; a Langevin thermostat is required for the Einstein crystal reference because a Nos\'e--Hoover chain thermostat can fail to ergodically sample independent harmonic oscillators. Each MD task in ice HTI spans 500~ps.

The following commands generate the three-step HTI tasks, run each step, and compute the Gibbs free energy:
\commandboxneedspace
\begin{center}
\begin{minipage}{0.5\textwidth}
\begin{grayverb}
\begin{spverbatim}
dpti hti_ice gen hti.ice.json \
  -o hti -s three-step
dpti hti_ice run hti \
  machine.della.cpu.json 00
dpti hti_ice run hti \
  machine.della.gpu.json 01
dpti hti_ice run hti \
  machine.della.gpu.json 02
cd hti
dpti hti_ice compute . \
  -t gibbs --npt ../npt
\end{spverbatim}
\end{grayverb}
\end{minipage}
\end{center}
\vspace{1em}

\begin{figure}[htbp]
\centering
\includegraphics[width=\textwidth]{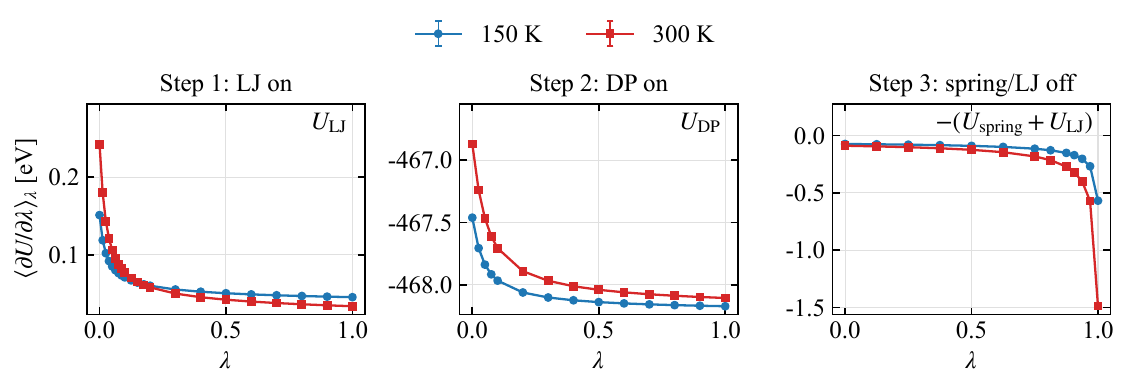}
\caption{HTI integrands $\langle\frac{\partial U}{\partial \lambda}\rangle_{\lambda}$ for ice Ih at 150 and 300~K along the three-step path from the Einstein crystal reference state to the target solid. The three panels correspond to switching on the soft-core LJ term, switching on the target DP potential, and removing the auxiliary spring and LJ terms, respectively.}
\label{fig:ice_hti_integrand}
\end{figure}

Figure~\ref{fig:ice_hti_integrand} shows the HTI integrands for ice Ih. The three panels correspond to switching on the auxiliary soft-core LJ interaction, switching on the target DP potential, and removing the Einstein spring together with the auxiliary LJ interaction.

The individual contributions to the final Gibbs free energy, as reported by the \texttt{dpti hti\_ice compute} post-processing command, are summarized in Table~\ref{tab:ice_hti_contributions}. The final results are the Gibbs free energy per H$_2$O molecule of ice Ih at 1~bar and 150 or 300~K.

\begin{table*}[t]
\centering
\caption{Contributions to the Gibbs free energy for the ice Ih HTI calculations. $A_0$ is the Frenkel-crystal reference free energy, $-TS_{\mathrm{conf}}$ is the proton-disorder correction, and $\Delta A_1$, $\Delta A_2$, and $\Delta A_3$ are the HTI contributions from the three steps. The Gibbs free energy is $G=A_0-TS_{\mathrm{conf}}+\Delta A_1+\Delta A_2+\Delta A_3+p_0\langle V\rangle$. The last seven columns are in $\mathrm{eV}$/H$_2$O. Numbers in parentheses denote the statistical uncertainty in the last digits.}
\label{tab:ice_hti_contributions}
\small
\setlength{\tabcolsep}{3pt}
\begin{tabular}{cccccccc}
\toprule
$T_0$ [K] & $A_0$ & $-TS_{\mathrm{conf}}$ & $\Delta A_1$ & $\Delta A_2$ & $\Delta A_3$ & $p_0\langle V\rangle$ & $G$ \\
\midrule
150 & $0.14640$ & $-0.00524$ & $0.055804(4)$ & $-468.09409(8)$ & $-0.11237(8)$ & $2.038(3)\times10^{-5}$ & $-468.0095(2)$ \\
300 & $0.13159$ & $-0.01048$ & $0.051968(6)$ & $-467.9566(1)$ & $-0.1806(3)$ & $2.109(7)\times10^{-5}$ & $-467.9641(4)$ \\
\bottomrule
\end{tabular}
\end{table*}

For water, the HTI tasks are generated from \texttt{hti.water.json}. This file specifies the settings for the liquid-water HTI workflow, including the $\lambda$ grids for the three HTI steps, the parameters of the soft-core LJ potential, the bond and angle spring constants, reference bond length and angle, and other MD settings. Each MD task in liquid-water HTI spans 500~ps.

The following commands generate the three-step HTI tasks, run each step, and compute the Gibbs free energy:
\commandboxneedspace
\begin{center}
\begin{minipage}{0.5\textwidth}
\begin{grayverb}
\begin{spverbatim}
dpti hti_water gen \
  hti.water.json -o hti
dpti hti_water run hti \
  machine.cpu.json 00 --no-dp
dpti hti_water run hti \
  machine.gpu.json 01
dpti hti_water run hti \
  machine.gpu.json 02
cd hti
dpti hti_water compute . \
  -t gibbs --npt ../npt
cd ..
\end{spverbatim}
\end{grayverb}
\end{minipage}
\end{center}
\vspace{1em}

\begin{figure}[htbp]
\centering
\includegraphics[width=\textwidth]{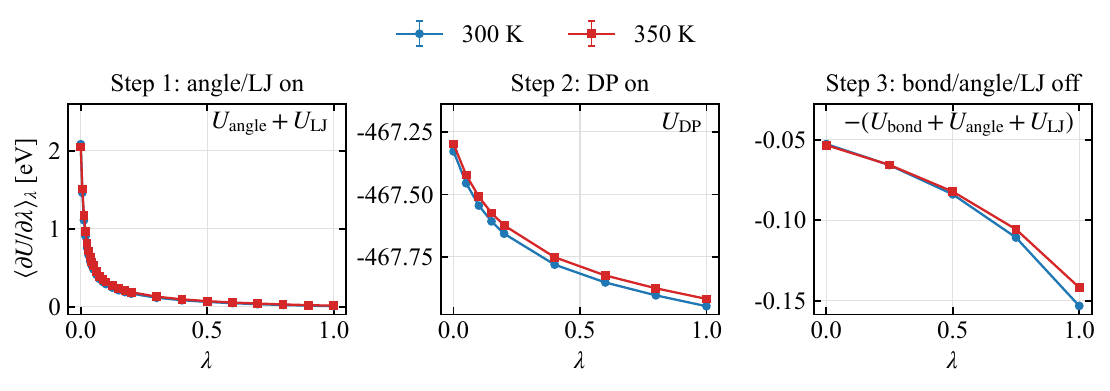}
\caption{HTI integrands $\langle\frac{\partial U}{\partial \lambda}\rangle_{\lambda}$ for liquid water at 300 and 350~K along the three-step path from the reference state of ideal molecules to the target liquid. The three panels correspond to switching on the angular and soft-core LJ terms, switching on the target DP potential, and removing the auxiliary bond, angle, and LJ terms, respectively.}
\label{fig:water_hti_integrand}
\end{figure}

Figure~\ref{fig:water_hti_integrand} shows the HTI integrands for liquid water. The three panels correspond to switching on the angular term together with the auxiliary soft-core LJ interaction, switching on the target DP potential, and removing the auxiliary bond, angle, and LJ interactions.

The individual contributions to the final Gibbs free energy, as reported by the \texttt{dpti hti\_water compute} post-processing command, are summarized in Table~\ref{tab:water_hti_contributions}. The final results are the Gibbs free energy per H$_2$O molecule of liquid water at 1~bar and 300 or 350~K.

\begin{table*}[t]
\centering
\caption{Contributions to the Gibbs free energy for the liquid-water HTI calculations. $A_0$ is the Helmholtz free energy of the ideal-molecule reference system, and $\Delta A_1$, $\Delta A_2$, and $\Delta A_3$ are the HTI contributions from the three steps. The Gibbs free energy is $G=A_0+\Delta A_1+\Delta A_2+\Delta A_3+p_0\langle V\rangle$. The last six columns are in $\mathrm{eV}$/H$_2$O. Numbers in parentheses denote the statistical uncertainty in the last digits.}
\label{tab:water_hti_contributions}
\small
\setlength{\tabcolsep}{3pt}
\begin{tabular}{ccccccc}
\toprule
$T_0$ [K] & $A_0$ & $\Delta A_1$ & $\Delta A_2$ & $\Delta A_3$ & $p_0\langle V\rangle$ & $G$ \\
\midrule
300 & $-0.23168$ & $0.13250(6)$ & $-467.7751(1)$ & $-0.08998(4)$ & $2.030(3)\times10^{-5}$ & $-467.9642(2)$ \\
350 & $-0.29646$ & $0.14698(6)$ & $-467.7452(1)$ & $-0.08706(4)$ & $2.068(1)\times10^{-5}$ & $-467.9817(2)$ \\
\bottomrule
\end{tabular}
\end{table*}

\subsubsection{Stage 4: Temperature Thermodynamic Integration}
After the HTI calculation provides the Gibbs free energy at $T_0$, TTI propagates it to other temperatures. The TTI commands are analogous for ice and liquid water, and each TTI MD simulation is run for 500~ps. Changing the HTI temperature $T_0$ is achieved by changing the HTI task directory specified after the \texttt{-H} option. The \texttt{ti\_water} command differs from the atomic TTI command \texttt{ti} only by normalizing the final Gibbs free energy per H$_2$O molecule instead of per atom, and is therefore used for both ice Ih and liquid water:
\commandboxneedspace
\begin{center}
\begin{minipage}{0.5\textwidth}
\begin{grayverb}
\begin{spverbatim}
dpti ti_water gen ti.t.json -o ti
dpti ti_water run ti \
  machine.gpu.json
dpti ti_water compute . -H ../hti
\end{spverbatim}
\end{grayverb}
\end{minipage}
\end{center}

\begin{figure}
\includegraphics[width=\textwidth]{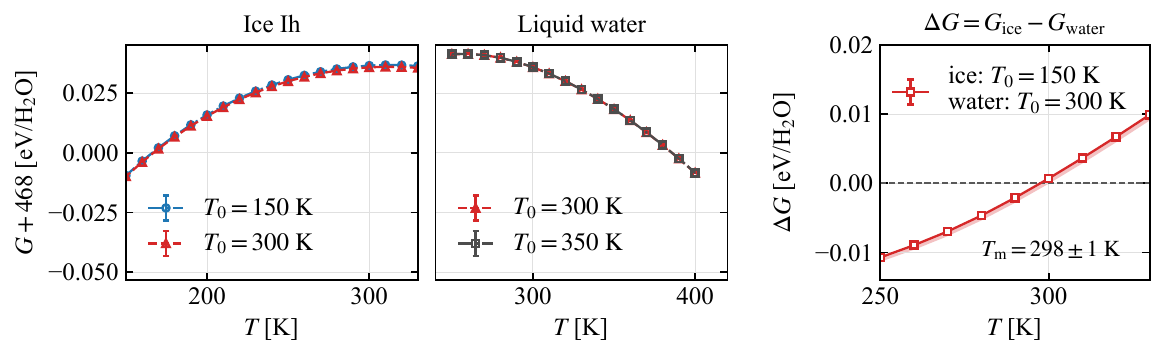}
\caption{Temperature thermodynamic integration of ice Ih and liquid water using the DP@revPBE-D3 model. The left and middle panels show the Gibbs free energy per water molecule propagated from different HTI temperatures $T_0$ for ice Ih and liquid water, respectively. The right panel shows the Gibbs free energy difference $\Delta G=G_{\mathrm{ice}}-G_{\mathrm{water}}$ for one representative pair of HTI temperatures; the shaded region indicates the range spanned by all four combinations of the two ice and two liquid-water HTI temperatures.}
\label{fig:tti_g_water_ice_deltag}
\end{figure}

Figure~\ref{fig:tti_g_water_ice_deltag} shows the TTI results obtained from the two independent HTI temperatures for ice Ih and liquid water. The agreement between the two propagated $G(T)$ curves for each phase provides a consistency check of the TI workflow and demonstrates the correctness of the HTI and TTI calculations. The right panel of Figure~\ref{fig:tti_g_water_ice_deltag} shows the Gibbs free energy difference $\Delta G=G_{\mathrm{ice}}-G_{\mathrm{water}}$ from one representative pair of HTI temperatures, yielding a melting temperature of $T_{\mathrm{m}}=298\pm1$~K at $p_0 = 1$~bar. 
The shaded region indicates the range spanned by all four combinations of the two ice and two liquid-water HTI temperatures, yielding a $T_{\mathrm{m}}$ range of 297.5 to 300.2~K. The small spread in $T_{\mathrm{m}}$ demonstrates the consistency of the HTI and TTI calculations. The resulting coexistence point, \(298\)~K and \(1\)~bar, is the starting point for the subsequent GDI calculation to trace the phase boundary.

\vspace{1em}
\subsubsection{Stage 5: Gibbs-Duhem Integration}
Starting from the coexistence point identified by TTI, GDI calculates the slope of the coexistence line using Eq.~\eqref{eq:gdi} and propagates the ice Ih--liquid water phase boundary. At each \((T,p)\) point, the enthalpy and volume differences between the two phases are evaluated using $NpT$ MD. Each phase-pair MD simulation is run for 100~ps. \texttt{dpti gdi} requires two JSON files to specify a GDI calculation: \texttt{in.gdi.json} defines the two phases, model, and MD settings, while \texttt{gdidata.json} defines the GDI settings such as the integration direction, initial coexistence point, and target points. For the ice Ih--liquid water boundary considered here, \(T\) changes only weakly with \(p\), so we use the \(p\) path, which uses pressure as the independent variable and propagates \(T(p)\).

The following command generates the GDI tasks, runs the MD simulations, computes the slope $\mathrm{d}T/\mathrm{d}p$ at each point, and propagates the phase boundary:
\commandboxneedspace
\begin{center}
\begin{minipage}{0.5\textwidth}
\begin{grayverb}
\begin{spverbatim}
dpti gdi in.gdi.json \
  machine.gpu.json -g gdidata.json
\end{spverbatim}
\end{grayverb}
\end{minipage}
\end{center}

\begin{figure}[htbp]
\centering
\includegraphics[width=\textwidth]{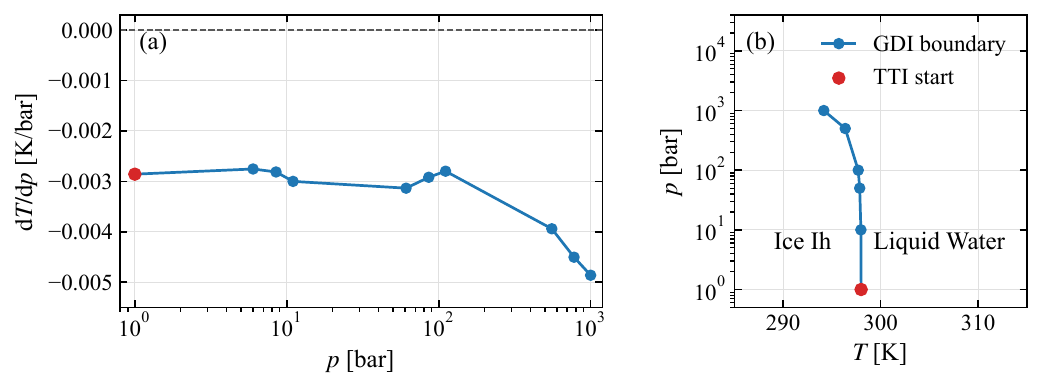}
\caption{Gibbs-Duhem integration of the ice Ih--liquid water phase boundary using the DP@revPBE-D3 model. (a) \(\mathrm{d}T/\mathrm{d}p\) evaluated from the MD simulations. (b) The propagated phase-boundary points, with the phase regions labeled. The red markers denote the coexistence point obtained from TTI and used to initialize the GDI calculation.}
\label{fig:water_gdi_phase_boundary}
\end{figure}

The evaluated \(\mathrm{d}T/\mathrm{d}p\) values are recorded in \texttt{dpdt.out}, and the propagated boundary points are recorded in \texttt{pb.out}. Figure~\ref{fig:water_gdi_phase_boundary}(a) shows the \(\mathrm{d}T/\mathrm{d}p\) values computed from the phase-pair MD simulations, and the negative values indicate that the melting temperature decreases with increasing pressure. Figure~\ref{fig:water_gdi_phase_boundary}(b) shows the resulting ice Ih--liquid water phase boundary up to 1000~bar. The calculated melting temperature at 1~bar is about 25~K higher than the experimental value of 273.15~K, which should be attributed mainly to the error of the underlying revPBE-D3 DFT functional.

Other water phases can be treated in the same way and combined with the ice Ih--liquid water boundary to construct a fuller water phase diagram~\cite{zhang_phase_2021}.

\section{Discussions}
TI is a powerful approach for computing free energies and phase boundaries, but reliable TI calculations require careful control of reference states, integration paths, MD sampling, numerical integration, and error estimation. The \texttt{dpti} workflow addresses these practical challenges by providing a systematic and automated framework for setting up, running, and post-processing TI calculations for material systems, thereby improving both reproducibility and robustness. \texttt{dpti} has already been used in several studies of phase diagrams driven by MLIPs. For single-element systems, it has been applied to tin~\cite{chen_modeling_2023,chen_deep_2026} and lithium~\cite{wang_data-driven_2023}. For mineral systems, it has been used to compute the order-disorder phase boundary of post-post-spinel Mg$_2$SiO$_4$ in super-Earth mantles~\cite{zheng_cation_2025} and the phase diagrams of the SiO$_2$, Al$_2$SiO$_5$, and Mg$_2$SiO$_4$ systems~\cite{zhong_general_2026}. For molecular systems, \texttt{dpti} has enabled phase diagram calculations for water~\cite{zhang_phase_2021,song_understanding_2026} and also supported the incorporation of nuclear quantum effects through path integral molecular dynamics~\cite{li_assessment_2025,li_ab_2025}. 

The design of a reversible integration path is crucial for successful TI calculations. In the current implementation, \texttt{dpti} provides several predefined HTI paths for the supported systems. Although flexible path designs are not currently supported, specialized TI paths for other systems can be incorporated by extending the existing implementation.

Another important limitation concerns the treatment of mixing free-energy contributions, including configurational entropy and excess mixing energy. As in other automated free-energy workflows~\cite{menon_automated_2021}, \texttt{dpti} does not automatically evaluate the free energy associated with substitutional disorder. In our recent work on sillimanite, the mixing free-energy of Al-Si atoms along the c-axis was instead calculated by combining cluster-expansion energy models with Monte Carlo sampling, which enables efficient evaluation of the configurational and excess mixing free energy at a substantially reduced computational cost~\cite{zhong_general_2026}. Incorporating such a mixing free-energy workflow into \texttt{dpti} is beyond the scope of the present study.

Molecular crystals are a natural target for extending the \texttt{dpti} workflow, but this extension is not trivial. Free energy calculations and comparisons of the relative stability of competing polymorphs of drug molecules are of great interest in the pharmaceutical industry and have motivated substantial research efforts~\cite{kapil_complete_2022}; related molecular-dynamics workflows have also been developed for computing solubilities of molecular and ionic crystals~\cite{reinhardt_streamlined_2023}. Compared with atomic solids, general molecular crystals require additional treatment of molecular orientations, intramolecular degrees of freedom, conformational changes, and the translational and rotational free-energy corrections needed to place different phases on a common reference scale. These ingredients make the design of reversible HTI paths more system-dependent. Free energy calculations for molecular liquids and solutions are also difficult. Even for water, one of the simplest molecular liquids, HTI requires multiple auxiliary interactions and carefully designed intermediate steps to ensure a reversible path. Moreover, empirical force fields often decompose the interaction into physically interpretable terms that can be turned on or off separately, whereas MLIPs usually represent the total potential energy through a flexible many-body model. As a result, it is less straightforward to isolate intramolecular, intermolecular, bonded, or nonbonded contributions when designing an HTI path. Further methodological developments will therefore be needed to make MLIP-based free energy calculations for complex molecular systems as routine as those for atomic materials.

\section{Conclusions}
We have presented \texttt{dpti}, a workflow for computing free energies and phase boundaries using thermodynamic integration. \texttt{dpti} provides an automated framework and software package for TI calculations, substantially reducing the tedious work of manually preparing, running, and post-processing large numbers of related MD tasks. We have demonstrated the use of \texttt{dpti} through two examples: the coesite--$\beta$-quartz--melt phase boundaries in silica and the ice Ih--liquid water phase boundary. These examples serve as pedagogical tutorials for applying \texttt{dpti} to atomic and molecular systems. We anticipate that \texttt{dpti} will be a useful tool for future studies of phase diagrams driven by MLIPs.

\numberedappendix{Center-of-Mass Correction in the Frenkel Reference}
\label{app:cm_correction}
In the Frenkel formulation, HTI is carried out in a CM-constrained subspace. The final target free energy can be written as
\begin{equation}
A_1
=
A_0^{\mathrm{CM}}
+
\Delta A_{0\rightarrow 1}^{\mathrm{CM}}
+
\left(A_1-A_1^{\mathrm{CM}}\right)
=
A_0^{\mathrm F}
+
\Delta A_{0\rightarrow 1}^{\mathrm{CM}},
\end{equation}
where $\Delta A_{0\rightarrow 1}^{\mathrm{CM}}=A_1^{\mathrm{CM}}-A_0^{\mathrm{CM}}$ is the HTI contribution evaluated in \texttt{dpti} under the CM constraint. Let \(\mathcal{P}\) denote the unconstrained momentum contribution. The CM-constrained Einstein crystal contains the corresponding constrained momentum contribution
\begin{equation}
\mathcal{P}_{\mathrm{CM}}
=
\frac{1}{h^{3(N-1)}}
\int
\exp\left[
-\beta\sum_i \frac{\mathbf{p}_i^2}{2m_i}
\right]
\delta\left(\sum_i \mathbf{p}_i\right)
\,\mathrm{d}\mathbf{p},
\end{equation}
which is not evaluated separately in \texttt{dpti}. The correction from the unconstrained Einstein reference in Eq.~\eqref{eq:einstein_reference} to the CM-constrained Einstein reference is
\begin{equation}
A_0^{\mathrm E}-A_0^{\mathrm{CM}}
=
-k_{\mathrm B}T\ln\frac{\mathcal{P}}{\mathcal{P}_{\mathrm{CM}}}
+
k_{\mathrm B}T
\left(
3\ln \Gamma+\frac{3}{2}\ln M
\right).
\end{equation}
The correction from the CM-constrained target solid to the unconstrained target solid is~\cite{zhang_phase_2021}
\begin{equation}
A_1-A_1^{\mathrm{CM}}
=
-k_{\mathrm B}T\ln\frac{\mathcal{P}}{\mathcal{P}_{\mathrm{CM}}}
+
k_{\mathrm B}T\ln\frac{N}{V}.
\end{equation}
The momentum terms cancel when the two corrections are combined, so the effective reference free energy used in \texttt{dpti} is
\begin{align}
A_0^{\mathrm F}
&=
A_0^{\mathrm E}
+
\left(A_0^{\mathrm{CM}}-A_0^{\mathrm E}\right)
+
\left(A_1-A_1^{\mathrm{CM}}\right) \nonumber\\
&=
A_0^{\mathrm E}
+
k_{\mathrm B}T
\left[
-3\ln \Gamma
-\frac{3}{2}\ln M
+\ln\frac{N}{V}
\right].
\end{align}
For a multicomponent atomic solid, this effective Frenkel reference free energy is
\begin{equation}
A_0^{\mathrm F}
=
3k_{\mathrm B}T
\sum_s N_s
\left(
\ln \Lambda_s
+
\ln \Gamma_s
\right)
+
k_{\mathrm B}T
\left[
-3\ln \Gamma
-\frac{3}{2}\ln M
+\ln\frac{N}{V}
\right],
\end{equation}
where $s$ indexes atomic species, $N_s$ is the number of atoms of species $s$, $N=\sum_s N_s$, $M=\sum_s N_s m_s$, $\Lambda_s=h/\sqrt{2\pi m_s k_{\mathrm B}T}$, and $\Gamma_s=\sqrt{k_s/(2\pi k_{\mathrm B}T)}$. The factor $\Gamma$ is computed from the mass-independent spring parameter used to construct the species-dependent spring constants.

\numberedappendix{Fixed-Particle Correction in the Vega Reference}
\label{app:vega_correction}
In the Vega, or Einstein molecule, formulation, one reference particle is fixed instead of the CM~\cite{vega_revisiting_2007}. If the fixed particle belongs to species $s_0$, the effective Vega reference free energy used in \texttt{dpti} is
\begin{equation}
A_0^{\mathrm V}
=
k_{\mathrm B}T
\left[
\sum_s 3N_s\ln \Lambda_s
+
3(N_{s_0}-1)\ln \Gamma_{s_0}
+
\sum_{s\ne s_0}3N_s\ln \Gamma_s
+
\ln\frac{N_{s_0}}{V}
\right].
\end{equation}
The fixed particle has no harmonic configurational contribution, giving the factor $N_{s_0}-1$ for species $s_0$, while the density term \(\ln(N_{s_0}/V)\) restores the translational degree of freedom removed by fixing the reference particle. Although the individual reference and correction terms differ from those in the Frenkel formulation, the same unconstrained target free energy is recovered when the chosen constraint is applied consistently along the HTI path.

\numberedappendix{Reference Free Energy for Ice}
\label{app:ice_reference}
In \texttt{dpti}, ice phases use the same Einstein-type atomic reference as other solid phases, with oxygen and hydrogen atoms harmonically restrained to their reference positions. For the Frenkel formulation, the effective reference free energy is obtained from Appendix~\ref{app:cm_correction} by taking \(s\in\{\mathrm O,\mathrm H\}\):
\begin{equation}
A_{0,\mathrm{ice}}^{\mathrm F}
=
3k_{\mathrm B}T
\sum_{s=\mathrm O,\mathrm H} N_s
\left(
\ln \Lambda_s
+
\ln \Gamma_s
\right)
+
k_{\mathrm B}T
\left[
-3\ln \Gamma
-\frac{3}{2}\ln M
+\ln\frac{N}{V}
\right]
-T S_{\mathrm{conf}},
\end{equation}
where \(N=N_{\mathrm O}+N_{\mathrm H}\), \(M=N_{\mathrm O}m_{\mathrm O}+N_{\mathrm H}m_{\mathrm H}\), \(\Lambda_s=h/\sqrt{2\pi m_s k_{\mathrm B}T}\), and \(\Gamma_s=\sqrt{k_s/(2\pi k_{\mathrm B}T)}\). The factor \(\Gamma\) is the corresponding mass-independent spring factor defined in Appendix~\ref{app:cm_correction}. The term \(-T S_{\mathrm{conf}}\) accounts for proton configurational disorder. For fully disordered ice phases, \texttt{dpti} uses Pauling's approximation~\cite{pauling_structure_1935},
\begin{equation}
\frac{S_{\mathrm{conf}}}{N_{\mathrm{H_2O}}}
\approx
k_{\mathrm B}\ln\frac{3}{2},
\end{equation}
where \(N_{\mathrm{H_2O}}=N_{\mathrm O}=N_{\mathrm H}/2\). For partially disordered ice III and ice V, the configurational entropy is estimated following Macdowell et al.~\cite{macdowell_combinatorial_2004}. If the fixed-particle Vega formulation is selected instead, the same configurational entropy correction is added to the corresponding \(A_0^{\mathrm V}\) in Appendix~\ref{app:vega_correction}.

\numberedappendix{Reference Free Energy for Liquid Water}
\label{app:water_reference}
The liquid-water reference used in \texttt{dpti} is an ideal gas of noninteracting water molecules. Each molecule \(\alpha\) contains one oxygen atom and two hydrogen atoms, and the hydrogen atoms are restrained to the oxygen atom by harmonic O--H springs:
\begin{equation}
H_0
=
\sum_i \frac{p_i^2}{2m_i}
+
\sum_{\alpha}
k_b\left(|\mathbf{r}^{\mathrm O}_{\alpha}-\mathbf{r}^{\mathrm H_1}_{\alpha}|-r_{\mathrm{OH},0}\right)^2
+
\sum_{\alpha}
k_b\left(|\mathbf{r}^{\mathrm O}_{\alpha}-\mathbf{r}^{\mathrm H_2}_{\alpha}|-r_{\mathrm{OH},0}\right)^2.
\end{equation}
Here \(k_b\) is the bond spring constant and \(r_{\mathrm{OH},0}\) is the equilibrium O--H bond length. The H--O--H angle is not constrained in this reference state. The configurational partition function contains a factor \(V^{N_{\mathrm O}}\) for the oxygen positions and a bond integral for each hydrogen atom,
\begin{equation}
Q_0^{\mathrm{mol}}
=
\left(\prod_{s=\mathrm O,\mathrm H} \Lambda_s^{-3N_s}\right)
\frac{V^{N_{\mathrm O}}}{N_{\mathrm O}!2^{N_{\mathrm O}}}
V_{\mathrm H}^{N_{\mathrm H}},
\end{equation}
where the factor \(2^{N_{\mathrm O}}\) accounts for the two equivalent hydrogen sites in each water molecule. The one-bond configurational integral is evaluated using the Gaussian expression used in \texttt{dpti},
\begin{align}
V_{\mathrm H}
&=
\int
\exp\left[-\beta k_b(r-r_{\mathrm{OH},0})^2\right]
\,\mathrm d\mathbf r \nonumber\\
&=
4\pi
\sqrt{\frac{\pi k_{\mathrm B}T}{k_b}}
\left(
r_{\mathrm{OH},0}^2
+
\frac{k_{\mathrm B}T}{2k_b}
\right).
\end{align}
Using Stirling's approximation for \(N_{\mathrm O}!\), the corresponding reference Helmholtz free energy can be written as~\cite{zhang_phase_2021}
\begin{equation}
A_0^{\mathrm{mol}}
=
k_{\mathrm B}T\sum_{s=\mathrm O,\mathrm H} 3N_s\ln\Lambda_s
+
N_{\mathrm O}k_{\mathrm B}T\ln \rho_{\mathrm O}
+
N_{\mathrm H}k_{\mathrm B}T \ln \left(\sqrt{2}/V_{\mathrm H}\right)
+
k_{\mathrm B}T\left(-N_{\mathrm O}+\frac{1}{2}\ln 2\pi N_{\mathrm O}\right),
\end{equation}
where \(\rho_{\mathrm O}=N_{\mathrm O}/V\). This is the analytical reference free energy used as the starting point for the liquid-water HTI path.

\begin{acknowledgement}

The authors thank Axel Gomez, Ping Tuo, and Xiaoyang Wang for helpful discussions. The authors also thank Roberto Car and Ryan Szukalo for their careful reading of the manuscript and helpful comments. Y.L. acknowledges support from the Computational Chemical Science Center: Chemistry in Solution and at Interfaces (CSI), funded by the U.S. Department of Energy under Award No. DE-SC0019394. The authors are pleased to acknowledge that the work reported on in this paper was performed using Princeton University's Research Computing resources.

\end{acknowledgement}

\section*{Data and Software Availability}

The data underlying this study are available in the published article, the Supporting Information, and at \url{https://github.com/Yi-FanLi/dpti_manuscript_examples}. The repository contains the input files, post-processed output data, and plotting or analysis files used for the silica and water examples. The \texttt{dpti} source code is available at \url{https://github.com/deepmodeling/dpti}.

\bibliography{clean}

\providecommand{\latin}[1]{#1}
\makeatletter
\providecommand{\doi}
  {\begingroup\let\do\@makeother\dospecials
  \catcode`\{=1 \catcode`\}=2 \doi@aux}
\providecommand{\doi@aux}[1]{\endgroup\texttt{#1}}
\makeatother
\providecommand*\mcitethebibliography{\thebibliography}
\csname @ifundefined\endcsname{endmcitethebibliography}  {\let\endmcitethebibliography\endthebibliography}{}
\begin{mcitethebibliography}{48}
\providecommand*\natexlab[1]{#1}
\providecommand*\mciteSetBstSublistMode[1]{}
\providecommand*\mciteSetBstMaxWidthForm[2]{}
\providecommand*\mciteBstWouldAddEndPuncttrue
  {\def\EndOfBibitem{\unskip.}}
\providecommand*\mciteBstWouldAddEndPunctfalse
  {\let\EndOfBibitem\relax}
\providecommand*\mciteSetBstMidEndSepPunct[3]{}
\providecommand*\mciteSetBstSublistLabelBeginEnd[3]{}
\providecommand*\EndOfBibitem{}
\mciteSetBstSublistMode{f}
\mciteSetBstMaxWidthForm{subitem}{(\alph{mcitesubitemcount})}
\mciteSetBstSublistLabelBeginEnd
  {\mcitemaxwidthsubitemform\space}
  {\relax}
  {\relax}

\bibitem[Chew and Reinhardt(2023)Chew, and Reinhardt]{chew_phase_2023}
Chew,~P.~Y.; Reinhardt,~A. Phase diagrams—{Why} they matter and how to predict them. \emph{J. Chem. Phys.} \textbf{2023}, \emph{158}\relax
\mciteBstWouldAddEndPuncttrue
\mciteSetBstMidEndSepPunct{\mcitedefaultmidpunct}
{\mcitedefaultendpunct}{\mcitedefaultseppunct}\relax
\EndOfBibitem
\bibitem[Frenkel and Ladd(1984)Frenkel, and Ladd]{frenkel_new_1984}
Frenkel,~D.; Ladd,~A. J.~C. New {Monte} {Carlo} method to compute the free energy of arbitrary solids. {Application} to the fcc and hcp phases of hard spheres. \emph{J. Chem. Phys.} \textbf{1984}, \emph{81}, 3188--3193\relax
\mciteBstWouldAddEndPuncttrue
\mciteSetBstMidEndSepPunct{\mcitedefaultmidpunct}
{\mcitedefaultendpunct}{\mcitedefaultseppunct}\relax
\EndOfBibitem
\bibitem[Polson \latin{et~al.}(2000)Polson, Trizac, Pronk, and Frenkel]{polson_finite-size_2000}
Polson,~J.~M.; Trizac,~E.; Pronk,~S.; Frenkel,~D. Finite-size corrections to the free energies of crystalline solids. \emph{The Journal of Chemical Physics} \textbf{2000}, \emph{112}, 5339--5342\relax
\mciteBstWouldAddEndPuncttrue
\mciteSetBstMidEndSepPunct{\mcitedefaultmidpunct}
{\mcitedefaultendpunct}{\mcitedefaultseppunct}\relax
\EndOfBibitem
\bibitem[Vega and Noya(2007)Vega, and Noya]{vega_revisiting_2007}
Vega,~C.; Noya,~E.~G. Revisiting the {Frenkel}-{Ladd} method to compute the free energy of solids: {The} {Einstein} molecule approach. \emph{J. Chem. Phys.} \textbf{2007}, \emph{127}, 154113\relax
\mciteBstWouldAddEndPuncttrue
\mciteSetBstMidEndSepPunct{\mcitedefaultmidpunct}
{\mcitedefaultendpunct}{\mcitedefaultseppunct}\relax
\EndOfBibitem
\bibitem[Vega \latin{et~al.}(2008)Vega, Sanz, Abascal, and Noya]{vega_determination_2008}
Vega,~C.; Sanz,~E.; Abascal,~J. L.~F.; Noya,~E.~G. Determination of phase diagrams via computer simulation: methodology and applications to water, electrolytes and proteins. \emph{J. Phys.: Condens. Matter} \textbf{2008}, \emph{20}, 153101\relax
\mciteBstWouldAddEndPuncttrue
\mciteSetBstMidEndSepPunct{\mcitedefaultmidpunct}
{\mcitedefaultendpunct}{\mcitedefaultseppunct}\relax
\EndOfBibitem
\bibitem[Sugino and Car(1995)Sugino, and Car]{sugino_ab_1995}
Sugino,~O.; Car,~R. \textit{{Ab} {Initio}} {Molecular} {Dynamics} {Study} of {First}-{Order} {Phase} {Transitions}: {Melting} of {Silicon}. \emph{Phys. Rev. Lett.} \textbf{1995}, \emph{74}, 1823--1826\relax
\mciteBstWouldAddEndPuncttrue
\mciteSetBstMidEndSepPunct{\mcitedefaultmidpunct}
{\mcitedefaultendpunct}{\mcitedefaultseppunct}\relax
\EndOfBibitem
\bibitem[Alfè \latin{et~al.}(1999)Alfè, Gillan, and Price]{alfe_melting_1999}
Alfè,~D.; Gillan,~M.~J.; Price,~G.~D. The melting curve of iron at the pressures of the {Earth}'s core from ab initio calculations. \emph{Nature} \textbf{1999}, \emph{401}, 462--464\relax
\mciteBstWouldAddEndPuncttrue
\mciteSetBstMidEndSepPunct{\mcitedefaultmidpunct}
{\mcitedefaultendpunct}{\mcitedefaultseppunct}\relax
\EndOfBibitem
\bibitem[Ghiringhelli \latin{et~al.}(2005)Ghiringhelli, Los, Meijer, Fasolino, and Frenkel]{ghiringhelli_modeling_2005}
Ghiringhelli,~L.~M.; Los,~J.~H.; Meijer,~E.~J.; Fasolino,~A.; Frenkel,~D. Modeling the {Phase} {Diagram} of {Carbon}. \emph{Phys. Rev. Lett.} \textbf{2005}, \emph{94}, 145701\relax
\mciteBstWouldAddEndPuncttrue
\mciteSetBstMidEndSepPunct{\mcitedefaultmidpunct}
{\mcitedefaultendpunct}{\mcitedefaultseppunct}\relax
\EndOfBibitem
\bibitem[Wang \latin{et~al.}(2005)Wang, Scandolo, and Car]{wang_carbon_2005}
Wang,~X.; Scandolo,~S.; Car,~R. Carbon {Phase} {Diagram} from \textit{{Ab} {Initio}} {Molecular} {Dynamics}. \emph{Phys. Rev. Lett.} \textbf{2005}, \emph{95}, 185701\relax
\mciteBstWouldAddEndPuncttrue
\mciteSetBstMidEndSepPunct{\mcitedefaultmidpunct}
{\mcitedefaultendpunct}{\mcitedefaultseppunct}\relax
\EndOfBibitem
\bibitem[Kaczmarski(2005)]{kaczmarski_phase_2005}
Kaczmarski,~M. Phase {Diagram} of {Silicon} from {Atomistic} {Simulations}. \emph{Phys. Rev. Lett.} \textbf{2005}, \emph{94}\relax
\mciteBstWouldAddEndPuncttrue
\mciteSetBstMidEndSepPunct{\mcitedefaultmidpunct}
{\mcitedefaultendpunct}{\mcitedefaultseppunct}\relax
\EndOfBibitem
\bibitem[Saika-Voivod(2004)]{saika-voivod_phase_2004}
Saika-Voivod,~I. Phase diagram of silica from computer simulation. \emph{Phys. Rev. E} \textbf{2004}, \emph{70}\relax
\mciteBstWouldAddEndPuncttrue
\mciteSetBstMidEndSepPunct{\mcitedefaultmidpunct}
{\mcitedefaultendpunct}{\mcitedefaultseppunct}\relax
\EndOfBibitem
\bibitem[Ford \latin{et~al.}(2007)Ford, Auerbach, and Monson]{ford_further_2007}
Ford,~M.~H.; Auerbach,~S.~M.; Monson,~P.~A. Further studies of a simple atomistic model of silica: {Thermodynamic} stability of zeolite frameworks as silica polymorphs. \emph{The Journal of Chemical Physics} \textbf{2007}, \emph{126}, 144701\relax
\mciteBstWouldAddEndPuncttrue
\mciteSetBstMidEndSepPunct{\mcitedefaultmidpunct}
{\mcitedefaultendpunct}{\mcitedefaultseppunct}\relax
\EndOfBibitem
\bibitem[Zhang \latin{et~al.}(2018)Zhang, Han, Wang, Car, and E]{zhang_deep_2018}
Zhang,~L.; Han,~J.; Wang,~H.; Car,~R.; E,~W. Deep {Potential} {Molecular} {Dynamics}: {A} {Scalable} {Model} with the {Accuracy} of {Quantum} {Mechanics}. \emph{Phys. Rev. Lett.} \textbf{2018}, \emph{120}, 143001\relax
\mciteBstWouldAddEndPuncttrue
\mciteSetBstMidEndSepPunct{\mcitedefaultmidpunct}
{\mcitedefaultendpunct}{\mcitedefaultseppunct}\relax
\EndOfBibitem
\bibitem[Zhang \latin{et~al.}(2018)Zhang, Han, Wang, Saidi, Car, and E]{zhang_end}
Zhang,~L.; Han,~J.; Wang,~H.; Saidi,~W.; Car,~R.; E,~W. End-to-end {Symmetry} {Preserving} {Inter}-atomic {Potential} {Energy} {Model} for {Finite} and {Extended} {Systems}. Advances in {Neural} {Information} {Processing} {Systems}. 2018\relax
\mciteBstWouldAddEndPuncttrue
\mciteSetBstMidEndSepPunct{\mcitedefaultmidpunct}
{\mcitedefaultendpunct}{\mcitedefaultseppunct}\relax
\EndOfBibitem
\bibitem[Wang \latin{et~al.}(2018)Wang, Zhang, Han, and E]{wang_deepmd-kit_2018}
Wang,~H.; Zhang,~L.; Han,~J.; E,~W. {DeePMD}-kit: {A} deep learning package for many-body potential energy representation and molecular dynamics. \emph{Computer Physics Communications} \textbf{2018}, \emph{228}, 178--184\relax
\mciteBstWouldAddEndPuncttrue
\mciteSetBstMidEndSepPunct{\mcitedefaultmidpunct}
{\mcitedefaultendpunct}{\mcitedefaultseppunct}\relax
\EndOfBibitem
\bibitem[Zeng \latin{et~al.}(2023)Zeng, Zhang, Lu, Mo, Li, Chen, Rynik, Huang, Li, Shi, Wang, Ye, Tuo, Yang, Ding, Li, Tisi, Zeng, Bao, Xia, Huang, Muraoka, Wang, Chang, Yuan, Bore, Cai, Lin, Wang, Xu, Zhu, Luo, Zhang, Goodall, Liang, Singh, Yao, Zhang, Wentzcovitch, Han, Liu, Jia, York, E, Car, Zhang, and Wang]{zeng_deepmd-kit_2023}
Zeng,~J. \latin{et~al.}  {DeePMD}-kit v2: {A} software package for deep potential models. \emph{The Journal of Chemical Physics} \textbf{2023}, \emph{159}, 054801\relax
\mciteBstWouldAddEndPuncttrue
\mciteSetBstMidEndSepPunct{\mcitedefaultmidpunct}
{\mcitedefaultendpunct}{\mcitedefaultseppunct}\relax
\EndOfBibitem
\bibitem[Zeng \latin{et~al.}(2025)Zeng, Zhang, Peng, Zhang, He, Wang, Liu, Bi, Li, Cai, Zhang, Du, Zhu, Mo, Huang, Zeng, Shi, Qin, Yu, Luo, Ding, Liu, Shi, Wang, Bore, Chang, Deng, Ding, Han, Jiang, Ke, Liu, Lu, Muraoka, Oliaei, Singh, Que, Xu, Xu, Zhuang, Dai, Giese, Jia, Xu, York, Zhang, and Wang]{zeng_deepmd-kit_2025}
Zeng,~J. \latin{et~al.}  {DeePMD}-kit v3: {A} {Multiple}-{Backend} {Framework} for {Machine} {Learning} {Potentials}. \emph{J. Chem. Theory Comput.} \textbf{2025}, \emph{21}, 4375--4385\relax
\mciteBstWouldAddEndPuncttrue
\mciteSetBstMidEndSepPunct{\mcitedefaultmidpunct}
{\mcitedefaultendpunct}{\mcitedefaultseppunct}\relax
\EndOfBibitem
\bibitem[De~Koning and Antonelli(1996)De~Koning, and Antonelli]{de_koning_einstein_1996}
De~Koning,~M.; Antonelli,~A. Einstein crystal as a reference system in free energy estimation using adiabatic switching. \emph{Phys. Rev. E} \textbf{1996}, \emph{53}, 465--474\relax
\mciteBstWouldAddEndPuncttrue
\mciteSetBstMidEndSepPunct{\mcitedefaultmidpunct}
{\mcitedefaultendpunct}{\mcitedefaultseppunct}\relax
\EndOfBibitem
\bibitem[Freitas \latin{et~al.}(2016)Freitas, Asta, and de~Koning]{freitas_nonequilibrium_2016}
Freitas,~R.; Asta,~M.; de~Koning,~M. Nonequilibrium free-energy calculation of solids using {LAMMPS}. \emph{Computational Materials Science} \textbf{2016}, \emph{112}, 333--341\relax
\mciteBstWouldAddEndPuncttrue
\mciteSetBstMidEndSepPunct{\mcitedefaultmidpunct}
{\mcitedefaultendpunct}{\mcitedefaultseppunct}\relax
\EndOfBibitem
\bibitem[Paula~Leite and De~Koning(2019)Paula~Leite, and De~Koning]{paula_leite_nonequilibrium_2019}
Paula~Leite,~R.; De~Koning,~M. Nonequilibrium free-energy calculations of fluids using {LAMMPS}. \emph{Computational Materials Science} \textbf{2019}, \emph{159}, 316--326\relax
\mciteBstWouldAddEndPuncttrue
\mciteSetBstMidEndSepPunct{\mcitedefaultmidpunct}
{\mcitedefaultendpunct}{\mcitedefaultseppunct}\relax
\EndOfBibitem
\bibitem[Plimpton(1995)]{plimpton_fast_1995}
Plimpton,~S. Fast {Parallel} {Algorithms} for {Short}-{Range} {Molecular} {Dynamics}. \emph{Journal of Computational Physics} \textbf{1995}, \emph{117}, 1--19\relax
\mciteBstWouldAddEndPuncttrue
\mciteSetBstMidEndSepPunct{\mcitedefaultmidpunct}
{\mcitedefaultendpunct}{\mcitedefaultseppunct}\relax
\EndOfBibitem
\bibitem[Thompson \latin{et~al.}(2022)Thompson, Aktulga, Berger, Bolintineanu, Brown, Crozier, in~'t Veld, Kohlmeyer, Moore, Nguyen, Shan, Stevens, Tranchida, Trott, and Plimpton]{thompson_lammps_2022}
Thompson,~A.~P.; Aktulga,~H.~M.; Berger,~R.; Bolintineanu,~D.~S.; Brown,~W.~M.; Crozier,~P.~S.; in~'t Veld,~P.~J.; Kohlmeyer,~A.; Moore,~S.~G.; Nguyen,~T.~D.; Shan,~R.; Stevens,~M.~J.; Tranchida,~J.; Trott,~C.; Plimpton,~S.~J. {LAMMPS} - a flexible simulation tool for particle-based materials modeling at the atomic, meso, and continuum scales. \emph{Computer Physics Communications} \textbf{2022}, \emph{271}, 108171\relax
\mciteBstWouldAddEndPuncttrue
\mciteSetBstMidEndSepPunct{\mcitedefaultmidpunct}
{\mcitedefaultendpunct}{\mcitedefaultseppunct}\relax
\EndOfBibitem
\bibitem[Menon \latin{et~al.}(2021)Menon, Lysogorskiy, Rogal, and Drautz]{menon_automated_2021}
Menon,~S.; Lysogorskiy,~Y.; Rogal,~J.; Drautz,~R. Automated free-energy calculation from atomistic simulations. \emph{Phys. Rev. Materials} \textbf{2021}, \emph{5}, 103801\relax
\mciteBstWouldAddEndPuncttrue
\mciteSetBstMidEndSepPunct{\mcitedefaultmidpunct}
{\mcitedefaultendpunct}{\mcitedefaultseppunct}\relax
\EndOfBibitem
\bibitem[Bonomi \latin{et~al.}(2009)Bonomi, Branduardi, Bussi, Camilloni, Provasi, Raiteri, Donadio, Marinelli, Pietrucci, Broglia, and Parrinello]{bonomi_plumed_2009}
Bonomi,~M.; Branduardi,~D.; Bussi,~G.; Camilloni,~C.; Provasi,~D.; Raiteri,~P.; Donadio,~D.; Marinelli,~F.; Pietrucci,~F.; Broglia,~R.~A.; Parrinello,~M. {PLUMED}: {A} portable plugin for free-energy calculations with molecular dynamics. \emph{Computer Physics Communications} \textbf{2009}, \emph{180}, 1961--1972\relax
\mciteBstWouldAddEndPuncttrue
\mciteSetBstMidEndSepPunct{\mcitedefaultmidpunct}
{\mcitedefaultendpunct}{\mcitedefaultseppunct}\relax
\EndOfBibitem
\bibitem[Tribello \latin{et~al.}(2014)Tribello, Bonomi, Branduardi, Camilloni, and Bussi]{tribello_plumed_2014}
Tribello,~G.~A.; Bonomi,~M.; Branduardi,~D.; Camilloni,~C.; Bussi,~G. {PLUMED} 2: {New} feathers for an old bird. \emph{Computer Physics Communications} \textbf{2014}, \emph{185}, 604--613\relax
\mciteBstWouldAddEndPuncttrue
\mciteSetBstMidEndSepPunct{\mcitedefaultmidpunct}
{\mcitedefaultendpunct}{\mcitedefaultseppunct}\relax
\EndOfBibitem
\bibitem[Alibay \latin{et~al.}(2025)Alibay, Gowers, Swenson, Henry, Ries, Baumann, Eastwood, Mitchell, Dotson, Horton, Thompson, and Travitz]{alibay_httpsgithubcomopenfreeenergyopenfe_2025}
Alibay,~I.; Gowers,~R.~J.; Swenson,~D.~W.; Henry,~M.~M.; Ries,~B.; Baumann,~H.~M.; Eastwood,~J. R.~B.; Mitchell,~A.; Dotson,~D.; Horton,~J.~T.; Thompson,~M.; Travitz,~A. https://github.com/{openfreeEnergy}/openfe. 2025; \url{https://github.com/openfreeEnergy/openfe}\relax
\mciteBstWouldAddEndPuncttrue
\mciteSetBstMidEndSepPunct{\mcitedefaultmidpunct}
{\mcitedefaultendpunct}{\mcitedefaultseppunct}\relax
\EndOfBibitem
\bibitem[Zhang \latin{et~al.}(2021)Zhang, Wang, Car, and E]{zhang_phase_2021}
Zhang,~L.; Wang,~H.; Car,~R.; E,~W. Phase {Diagram} of a {Deep} {Potential} {Water} {Model}. \emph{Phys. Rev. Lett.} \textbf{2021}, \emph{126}, 236001\relax
\mciteBstWouldAddEndPuncttrue
\mciteSetBstMidEndSepPunct{\mcitedefaultmidpunct}
{\mcitedefaultendpunct}{\mcitedefaultseppunct}\relax
\EndOfBibitem
\bibitem[De~Witte \latin{et~al.}(2026)De~Witte, Braeckevelt, Bocus, Vandenhaute, and Van~Speybroeck]{de_witte_novel_2026}
De~Witte,~K. L.~K.; Braeckevelt,~T.; Bocus,~M.; Vandenhaute,~S.; Van~Speybroeck,~V. A {Novel} {NPT} {Thermodynamic} {Integration} {Scheme} to {Derive} {Rigorous} {Gibbs} {Free} {Energies} for {Crystalline} {Solids}. \emph{J. Chem. Theory Comput.} \textbf{2026}, \relax
\mciteBstWouldAddEndPunctfalse
\mciteSetBstMidEndSepPunct{\mcitedefaultmidpunct}
{}{\mcitedefaultseppunct}\relax
\EndOfBibitem
\bibitem[Kofke(1993)]{kofke_direct_1993}
Kofke,~D.~A. Direct evaluation of phase coexistence by molecular simulation via integration along the saturation line. \emph{The Journal of Chemical Physics} \textbf{1993}, \emph{98}, 4149--4162\relax
\mciteBstWouldAddEndPuncttrue
\mciteSetBstMidEndSepPunct{\mcitedefaultmidpunct}
{\mcitedefaultendpunct}{\mcitedefaultseppunct}\relax
\EndOfBibitem
\bibitem[Pauling(1935)]{pauling_structure_1935}
Pauling,~L. The {Structure} and {Entropy} of {Ice} and of {Other} {Crystals} with {Some} {Randomness} of {Atomic} {Arrangement}. \emph{J. Am. Chem. Soc.} \textbf{1935}, \emph{57}, 2680--2684\relax
\mciteBstWouldAddEndPuncttrue
\mciteSetBstMidEndSepPunct{\mcitedefaultmidpunct}
{\mcitedefaultendpunct}{\mcitedefaultseppunct}\relax
\EndOfBibitem
\bibitem[MacDowell \latin{et~al.}(2004)MacDowell, Sanz, Vega, and Abascal]{macdowell_combinatorial_2004}
MacDowell,~L.~G.; Sanz,~E.; Vega,~C.; Abascal,~J. L.~F. Combinatorial entropy and phase diagram of partially ordered ice phases. \emph{The Journal of Chemical Physics} \textbf{2004}, \emph{121}, 10145--10158\relax
\mciteBstWouldAddEndPuncttrue
\mciteSetBstMidEndSepPunct{\mcitedefaultmidpunct}
{\mcitedefaultendpunct}{\mcitedefaultseppunct}\relax
\EndOfBibitem
\bibitem[noa(2024)]{noauthor_httpsgithubcomdeepmodelingdpti_2024}
https://github.com/deepmodeling/dpti. 2024; \url{https://github.com/deepmodeling/dpti}\relax
\mciteBstWouldAddEndPuncttrue
\mciteSetBstMidEndSepPunct{\mcitedefaultmidpunct}
{\mcitedefaultendpunct}{\mcitedefaultseppunct}\relax
\EndOfBibitem
\bibitem[Li()]{li_httpsgithubcomyi-fanlidpti_manuscript_examples_nodate}
Li,~Y. https://github.com/{Yi}-{FanLi}/dpti\_manuscript\_examples. \url{https://github.com/Yi-FanLi/dpti_manuscript_examples}\relax
\mciteBstWouldAddEndPuncttrue
\mciteSetBstMidEndSepPunct{\mcitedefaultmidpunct}
{\mcitedefaultendpunct}{\mcitedefaultseppunct}\relax
\EndOfBibitem
\bibitem[Zhong \latin{et~al.}(2026)Zhong, Li, and John]{zhong_general_2026}
Zhong,~X.; Li,~Y.; John,~T. A general purposed machine learning interatomic potential for {Mg}-{Al}-{Si}-{O} system suitable for {Earth} materials at high pressure and temperature conditions. \emph{npj Comput Mater} \textbf{2026}, \relax
\mciteBstWouldAddEndPunctfalse
\mciteSetBstMidEndSepPunct{\mcitedefaultmidpunct}
{}{\mcitedefaultseppunct}\relax
\EndOfBibitem
\bibitem[Furness \latin{et~al.}(2020)Furness, Kaplan, Ning, Perdew, and Sun]{furness_accurate_2020}
Furness,~J.~W.; Kaplan,~A.~D.; Ning,~J.; Perdew,~J.~P.; Sun,~J. Accurate and {Numerically} {Efficient} {r2SCAN} {Meta}-{Generalized} {Gradient} {Approximation}. \emph{J. Phys. Chem. Lett.} \textbf{2020}, \emph{11}, 8208--8215\relax
\mciteBstWouldAddEndPuncttrue
\mciteSetBstMidEndSepPunct{\mcitedefaultmidpunct}
{\mcitedefaultendpunct}{\mcitedefaultseppunct}\relax
\EndOfBibitem
\bibitem[Yuan \latin{et~al.}(2025)Yuan, Ding, Liu, Cao, Fan, Nguyen, Zhang, Wang, Chen, Huang, Wen, Liu, Li, Zhuang, Yu, Tuo, Zhang, Wang, Zhang, Wang, and Zeng]{yuan_dpdispatcher_2025}
Yuan,~F. \latin{et~al.}  {DPDispatcher}: {Scalable} {HPC} {Task} {Scheduling} for {AI}-{Driven} {Science}. \emph{J. Chem. Inf. Model.} \textbf{2025}, \emph{65}, 12155--12160\relax
\mciteBstWouldAddEndPuncttrue
\mciteSetBstMidEndSepPunct{\mcitedefaultmidpunct}
{\mcitedefaultendpunct}{\mcitedefaultseppunct}\relax
\EndOfBibitem
\bibitem[Zhang \latin{et~al.}(2020)Zhang, Wang, Chen, Zeng, Zhang, Wang, and E]{zhang_dp-gen_2020}
Zhang,~Y.; Wang,~H.; Chen,~W.; Zeng,~J.; Zhang,~L.; Wang,~H.; E,~W. {DP}-{GEN}: {A} concurrent learning platform for the generation of reliable deep learning based potential energy models. \emph{Computer Physics Communications} \textbf{2020}, \emph{253}, 107206\relax
\mciteBstWouldAddEndPuncttrue
\mciteSetBstMidEndSepPunct{\mcitedefaultmidpunct}
{\mcitedefaultendpunct}{\mcitedefaultseppunct}\relax
\EndOfBibitem
\bibitem[Swamy \latin{et~al.}(1994)Swamy, Saxena, Sundman, and Zhang]{swamy_thermodynamic_1994}
Swamy,~V.; Saxena,~S.~K.; Sundman,~B.; Zhang,~J. A thermodynamic assessment of silica phase diagram. \emph{Journal of Geophysical Research: Solid Earth} \textbf{1994}, \emph{99}, 11787--11794\relax
\mciteBstWouldAddEndPuncttrue
\mciteSetBstMidEndSepPunct{\mcitedefaultmidpunct}
{\mcitedefaultendpunct}{\mcitedefaultseppunct}\relax
\EndOfBibitem
\bibitem[Li \latin{et~al.}(2025)Li, Yang, Zhang, Gomez, Xie, Chen, Piaggi, and Car]{li_ab_2025}
Li,~Y.; Yang,~B.; Zhang,~C.; Gomez,~A.; Xie,~P.; Chen,~Y.; Piaggi,~P.~M.; Car,~R. Ab {Initio} {Melting} {Properties} of {Water} and {Ice} from {Machine} {Learning} {Potentials}. 2025; \url{http://arxiv.org/abs/2512.23939}\relax
\mciteBstWouldAddEndPuncttrue
\mciteSetBstMidEndSepPunct{\mcitedefaultmidpunct}
{\mcitedefaultendpunct}{\mcitedefaultseppunct}\relax
\EndOfBibitem
\bibitem[Chen \latin{et~al.}(2023)Chen, Yuan, Liu, Geng, Zhang, Wang, and Chen]{chen_modeling_2023}
Chen,~T.; Yuan,~F.; Liu,~J.; Geng,~H.; Zhang,~L.; Wang,~H.; Chen,~M. Modeling the high-pressure solid and liquid phases of tin from deep potentials with ab initio accuracy. \emph{Phys. Rev. Mater.} \textbf{2023}, \emph{7}, 053603\relax
\mciteBstWouldAddEndPuncttrue
\mciteSetBstMidEndSepPunct{\mcitedefaultmidpunct}
{\mcitedefaultendpunct}{\mcitedefaultseppunct}\relax
\EndOfBibitem
\bibitem[Chen \latin{et~al.}(2026)Chen, Wang, Li, Chen, and Wang]{chen_deep_2026}
Chen,~Y.; Wang,~X.; Li,~W.; Chen,~M.; Wang,~H. Deep learning potential for accurate shock response simulations in tin. \emph{Phys. Rev. Materials} \textbf{2026}, \emph{10}, 023605\relax
\mciteBstWouldAddEndPuncttrue
\mciteSetBstMidEndSepPunct{\mcitedefaultmidpunct}
{\mcitedefaultendpunct}{\mcitedefaultseppunct}\relax
\EndOfBibitem
\bibitem[Wang \latin{et~al.}(2023)Wang, Wang, Gao, Zhang, Lv, Wang, Liu, Wang, and Ma]{wang_data-driven_2023}
Wang,~X.; Wang,~Z.; Gao,~P.; Zhang,~C.; Lv,~J.; Wang,~H.; Liu,~H.; Wang,~Y.; Ma,~Y. Data-driven prediction of complex crystal structures of dense lithium. \emph{Nat Commun} \textbf{2023}, \emph{14}, 2924\relax
\mciteBstWouldAddEndPuncttrue
\mciteSetBstMidEndSepPunct{\mcitedefaultmidpunct}
{\mcitedefaultendpunct}{\mcitedefaultseppunct}\relax
\EndOfBibitem
\bibitem[Zheng \latin{et~al.}(2025)Zheng, Li, Peng, Gong, Liu, and Deng]{zheng_cation_2025}
Zheng,~D.; Li,~Y.; Peng,~Y.; Gong,~R.; Liu,~Z.-K.; Deng,~J. Cation {Disorder} of \{\vphantom{\}}{\textbackslash}{textMg}\_{\textbackslash}mathbf2{\textbackslash}{textSiO}\_{\textbackslash}mathbf4{\textbackslash} in {Super}-{Earth} {Mantles}. \emph{Geophysical Research Letters} \textbf{2025}, \emph{52}, e2025GL118419\relax
\mciteBstWouldAddEndPuncttrue
\mciteSetBstMidEndSepPunct{\mcitedefaultmidpunct}
{\mcitedefaultendpunct}{\mcitedefaultseppunct}\relax
\EndOfBibitem
\bibitem[Song \latin{et~al.}(2026)Song, Liu, Zhang, Li, Santra, Chen, Klein, and Wu]{song_understanding_2026}
Song,~Y.; Liu,~R.; Zhang,~C.; Li,~Y.; Santra,~B.; Chen,~M.; Klein,~M.~L.; Wu,~X. Understanding the density maximum of water with machine-learned potentials. \emph{Science Advances} \textbf{2026}, \emph{12}, eaec6748\relax
\mciteBstWouldAddEndPuncttrue
\mciteSetBstMidEndSepPunct{\mcitedefaultmidpunct}
{\mcitedefaultendpunct}{\mcitedefaultseppunct}\relax
\EndOfBibitem
\bibitem[Li \latin{et~al.}(2025)Li, Yang, Zhang, Gomez, Xie, Chen, Piaggi, and Car]{li_assessment_2025}
Li,~Y.; Yang,~B.; Zhang,~C.; Gomez,~A.; Xie,~P.; Chen,~Y.; Piaggi,~P.~M.; Car,~R. Assessment of {First}-{Principles} {Methods} in {Modeling} the {Melting} {Properties} of {Water}. 2025; \url{http://arxiv.org/abs/2512.23940}\relax
\mciteBstWouldAddEndPuncttrue
\mciteSetBstMidEndSepPunct{\mcitedefaultmidpunct}
{\mcitedefaultendpunct}{\mcitedefaultseppunct}\relax
\EndOfBibitem
\bibitem[Kapil and Engel(2022)Kapil, and Engel]{kapil_complete_2022}
Kapil,~V.; Engel,~E.~A. A complete description of thermodynamic stabilities of molecular crystals. \emph{Proceedings of the National Academy of Sciences} \textbf{2022}, \emph{119}, e2111769119\relax
\mciteBstWouldAddEndPuncttrue
\mciteSetBstMidEndSepPunct{\mcitedefaultmidpunct}
{\mcitedefaultendpunct}{\mcitedefaultseppunct}\relax
\EndOfBibitem
\bibitem[Reinhardt \latin{et~al.}(2023)Reinhardt, Chew, and Cheng]{reinhardt_streamlined_2023}
Reinhardt,~A.; Chew,~P.~Y.; Cheng,~B. A streamlined molecular-dynamics workflow for computing solubilities of molecular and ionic crystals. \emph{J. Chem. Phys.} \textbf{2023}, \emph{159}\relax
\mciteBstWouldAddEndPuncttrue
\mciteSetBstMidEndSepPunct{\mcitedefaultmidpunct}
{\mcitedefaultendpunct}{\mcitedefaultseppunct}\relax
\EndOfBibitem
\end{mcitethebibliography}

\end{document}


\maketitle

\section{JSON Input Parameters}

\texttt{dpti} uses JSON files to define the molecular dynamics tasks and thermodynamic integration paths for each workflow stage. We use stage numbers consistent with the main text: 1 for $NpT$, 2 for $NVT$, 3 for HTI, 4 for TTI/pTI, and 5 for GDI. The label M denotes the optional \texttt{machine.json} file used for task submission. Table~\ref{tab:json_parameters} summarizes the commonly used parameters in these input files. The examples in the main text use these parameters with phase-specific values.

\begin{longtable}{L{0.10\textwidth}L{0.28\textwidth}L{0.54\textwidth}}
\caption{Common JSON parameters used by \texttt{dpti}.}
\label{tab:json_parameters}\\
\toprule
Stage & Parameter & Meaning \\
\midrule
\endfirsthead
\toprule
Stage & Parameter & Meaning \\
\midrule
\endhead
\bottomrule
\endfoot

1--5 &
\texttt{equi\_conf} &
Path to the initial LAMMPS data file. \\

1--5 &
\texttt{model} &
Path to the MLIP model file used by LAMMPS, for example a Deep Potential \texttt{graph.pb} file. \\

1--5 &
\texttt{mass\_map} &
List of atomic masses, ordered consistently with the atom types in the LAMMPS data file and the MLIP model. \\

1--5 &
\texttt{nsteps} &
Number of MD timesteps for each task. \\

1--5 &
\texttt{timestep} &
MD timestep, in picoseconds for the LAMMPS \texttt{metal} unit system used in the examples. \\

1--5 &
\texttt{ens} &
Simulation ensemble and barostat mode, such as \texttt{nvt}, \texttt{npt} (equivalent to \texttt{npt-iso}), \texttt{npt-iso}, \texttt{npt-aniso}, \texttt{npt-xy}, or \texttt{npt-tri}. \\

1--5 &
\texttt{temp}, \texttt{pres} &
Target temperature and pressure for MD simulations. Pressure is specified in bar for the LAMMPS \texttt{metal} unit system. \\

1--5 &
\texttt{tau\_t}, \texttt{tau\_p} &
Thermostat and barostat damping times. \\

1--5 &
\texttt{thermo\_freq}, \texttt{dump\_freq} &
Output frequencies for thermodynamic quantities and dumped configurations. \\

1--5 &
\texttt{stat\_skip} &
Number of initial samples discarded before computing statistical averages. \\

1--5 &
\texttt{stat\_bsize} &
Block size used for block averaging and statistical-error estimation. \\

1--5 &
\texttt{if\_meam}, \texttt{meam\_model} &
Options for using a MEAM potential instead of an MLIP model. \\

2 &
\texttt{if\_dump\_avg\_posi} &
Whether to write trajectory-averaged atomic positions. This option is useful for crystalline $NVT$ tasks that prepare reference positions for solid-state HTI, but should not be used for diffusive liquid configurations. \\

3, 4 &
\texttt{copies} &
Replication factors used to build the simulation cell in HTI and TTI/pTI task generation. When present, this key is also used to normalize extensive quantities by the replicated system size during post-processing. \\

3 &
\texttt{spring\_k} &
Mass-normalized spring constant $\kappa$ of the harmonic reference system. For each species $s$, the spring constant is $k_s=m_s\kappa$.\\

3 &
\texttt{crystal} &
Choice of solid reference formulation, such as \texttt{frenkel} or \texttt{vega}. \\

3 &
\texttt{langevin} &
Whether to use a Langevin thermostat. If \texttt{false}, Nos\'e--Hoover-chain thermostat is used. This is often set to \texttt{true} for nearly harmonic solid reference systems to avoid ergodicity problems of Nos\'e--Hoover-chain thermostats. \\

3 &
\texttt{lambda\_lj\_on}, \texttt{lambda\_deep\_on}, \texttt{lambda\_spring\_off} &
$\lambda$ grids for multi-step solid or ice HTI paths that switch on auxiliary LJ interactions, switch on the target MLIP potential, and switch off harmonic springs. \\

3 &
\texttt{lambda\_soft\_on}, \texttt{lambda\_soft\_off} &
$\lambda$ grids for liquid HTI steps that switch on and switch off the auxiliary soft-core LJ interaction. \\

3 &
\texttt{lambda\_angle\_on}, \texttt{lambda\_bond\_\allowbreak angle\_\allowbreak off} &
$\lambda$ grids for liquid-water HTI steps that introduce the angular restraint and remove the auxiliary molecular reference interactions. \\

3 &
\texttt{protect\_eps} &
Small numerical offset used to avoid singular behavior at endpoints of soft-core or switching functions. \\

3 &
\texttt{soft\_param} &
Parameters of the auxiliary soft-core LJ potential, including pairwise \texttt{sigma}, \texttt{epsilon}, and \texttt{activation} values when applicable, as well as \texttt{n}, \texttt{alpha\_lj}, and \texttt{rcut}. \\

3 &
\texttt{bond\_param} &
Parameters of the ideal-molecule water reference, including O--H bond spring constant and equilibrium length, and H--O--H angle spring constant and equilibrium angle. \\

3 &
\texttt{is\_water} &
Flag indicating that the water-specific HTI workflow and molecular reference should be used. \\

4 &
\texttt{path} &
Thermodynamic integration path. \texttt{t} denotes TTI along an isobar, while \texttt{p} denotes pTI along an isotherm. \\

4 &
\texttt{temp\_seq}, \texttt{pres\_seq} &
Temperature or pressure grid for TTI or pTI. A string such as \texttt{"200:1800:100"} denotes a regularly spaced sequence. \\

5 &
\texttt{direction} &
Independent variable used in GDI. \texttt{p} integrates $T(p)$, while \texttt{t} integrates $p(T)$. \\

5 &
\texttt{begin}, \texttt{end} &
Initial and final values of the independent variable for GDI. \\

5 &
\texttt{initial\_value} &
Initial value of the dependent variable at the starting coexistence point. \\

5 &
\texttt{abs\_tol}, \texttt{rel\_tol} &
Absolute and relative tolerances used by the adaptive ODE solver in GDI. \\

5 &
\texttt{phase\_i}, \texttt{phase\_ii} &
Definitions of the two coexisting phases, including phase names, initial configurations, and ensembles used for the MD simulations that evaluate the GDI slope. \\

M &
\texttt{machine} &
Machine and context settings used by \texttt{DPDispatcher}, such as batch-system type, context type, and local or remote working directory. \\

M &
\texttt{resources} &
Computational resources requested for each submission, such as number of nodes, CPUs or GPUs per node, queue name, and group size. \\

\end{longtable}

\section{Silica HTI Spring-Constant Check}

\begin{figure}[htbp]
\centering
\includegraphics[width=0.55\textwidth]{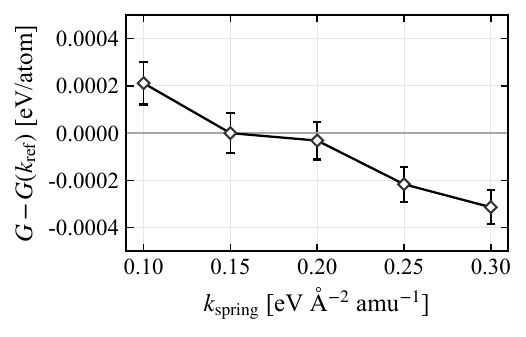}
\caption{Dependence of the coesite HTI Gibbs free energy on the mass-normalized spring constant at \(T=1600\)~K and \(p=5\)~GPa. Relative to \(k_{\mathrm{ref}}=0.15~\mathrm{eV}~\mbox{\AA}^{-2}~\mathrm{amu}^{-1}\), the deviations remain below 0.4~meV/atom.}
\label{fig:si_silica_hti_spring_k}
\end{figure}

Figure~\ref{fig:si_silica_hti_spring_k} shows the dependence of the coesite HTI Gibbs free energy on the mass-normalized spring constant used for the harmonic reference state. The relative Gibbs free energy is defined as \(G-G(k_{\mathrm{ref}})\), with \(T=1600\)~K and \(p=5\)~GPa. Relative to \(k_{\mathrm{ref}}=0.15~\mathrm{eV}~\mbox{\AA}^{-2}~\mathrm{amu}^{-1}\).

\section{Silica pTI Consistency Check}

\begin{figure}[htbp]
\centering
\includegraphics[width=\textwidth]{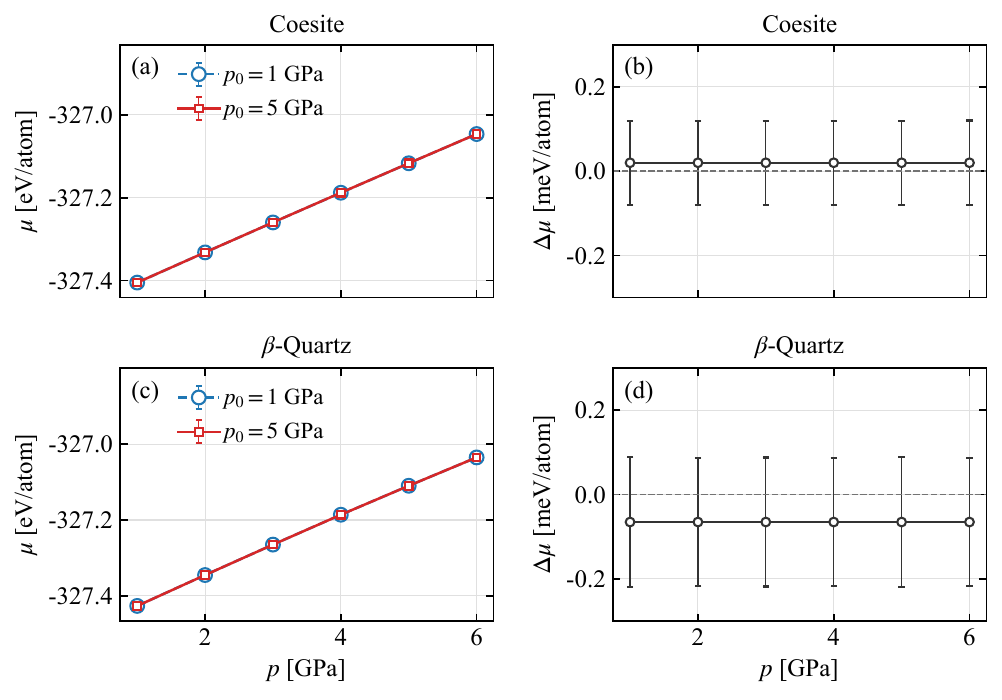}
\caption{Pressure thermodynamic integration consistency check for crystalline silica at $T=1600$~K. The left panels show the chemical potentials of coesite and $\beta$-quartz propagated from the two anchor pressures, $p_0=1$ and 5~GPa. The right panels show the corresponding differences, $\Delta\mu=\mu_{p_0=5\,\mathrm{GPa}}-\mu_{p_0=1\,\mathrm{GPa}}$, between the two propagated curves.}
\label{fig:si_silica_pti_anchor_consistency}
\end{figure}

Figure~\ref{fig:si_silica_pti_anchor_consistency} compares the pTI chemical potentials obtained from the two solid-phase anchor pressures and the corresponding anchor-to-anchor differences. Here, $\Delta\mu$ is defined as the chemical potential propagated from the higher-pressure HTI anchor minus that propagated from the lower-pressure HTI anchor, i.e., $\mu_{p_0=5\,\mathrm{GPa}}-\mu_{p_0=1\,\mathrm{GPa}}$.

For both crystalline phases, the two pTI curves propagated from the independent anchor pressures agree within their statistical uncertainties across the pressure range considered. This agreement supports the internal consistency of the HTI anchor free energies and the subsequent pTI propagation used to locate the coesite--$\beta$-quartz transition pressure.

\section{Silica TTI Consistency Checks}

\begin{figure}[htbp]
\centering
\includegraphics[width=\textwidth]{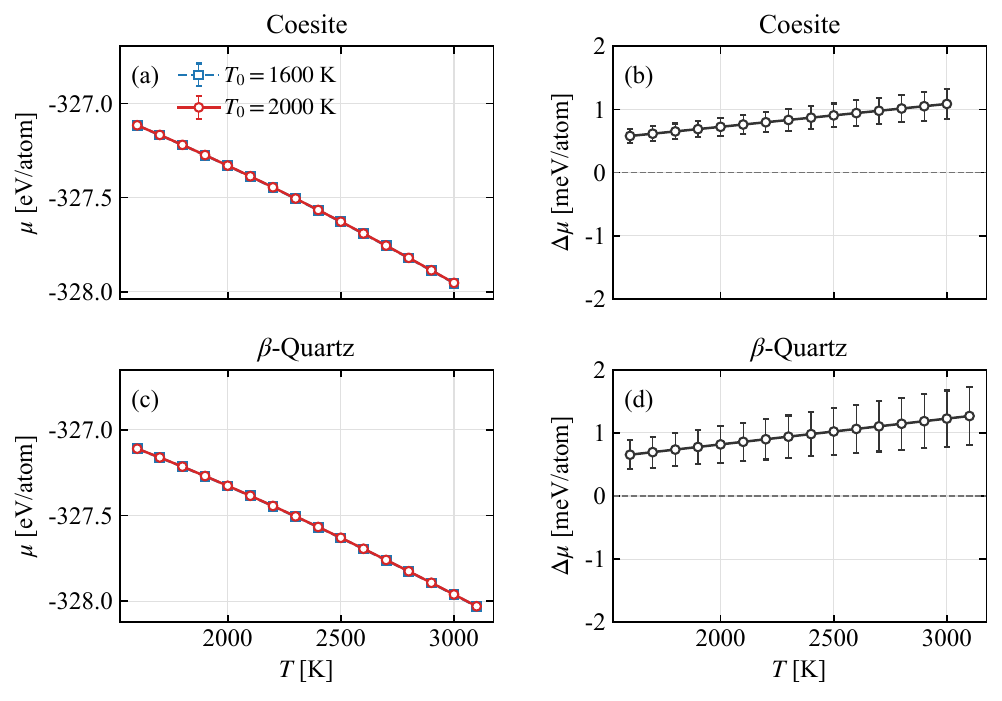}
\caption{Temperature thermodynamic integration consistency check for crystalline silica at $p=5$~GPa. The left panels show the chemical potentials of coesite and $\beta$-quartz propagated from the two HTI anchor temperatures, $T_0=1600$ and 2000~K. The right panels show the corresponding differences, $\Delta\mu=\mu_{T_0=2000\,\mathrm{K}}-\mu_{T_0=1600\,\mathrm{K}}$, between the two propagated curves.}
\label{fig:si_silica_solid_tti_anchor_consistency}
\end{figure}

\begin{figure}[htbp]
\centering
\includegraphics[width=\textwidth]{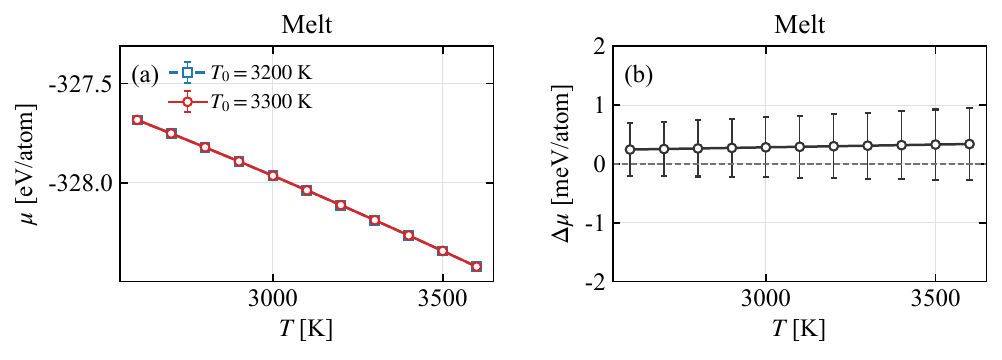}
\caption{Temperature thermodynamic integration consistency check for silica melt at $p=5$~GPa. The left panel shows the chemical potentials propagated from the two HTI anchor temperatures used in the main text, $T_0=3200$ and 3300~K. The right panel shows the difference, $\Delta\mu=\mu_{T_0=3300\,\mathrm{K}}-\mu_{T_0=3200\,\mathrm{K}}$, between the two propagated curves.}
\label{fig:si_silica_melt_tti_anchor_consistency}
\end{figure}

Figures~\ref{fig:si_silica_solid_tti_anchor_consistency} and \ref{fig:si_silica_melt_tti_anchor_consistency} show analogous TTI consistency checks for the crystalline phases and the silica melt using the anchor temperatures discussed in the main text. In these comparisons, $\Delta\mu$ is defined as the chemical potential propagated from the higher-temperature HTI anchor minus that propagated from the lower-temperature HTI anchor, i.e., $\mu_{T_0=2000\,\mathrm{K}}-\mu_{T_0=1600\,\mathrm{K}}$ for the crystalline phases and $\mu_{T_0=3300\,\mathrm{K}}-\mu_{T_0=3200\,\mathrm{K}}$ for the melt.

For both crystalline phases, Fig.~\ref{fig:si_silica_solid_tti_anchor_consistency} shows that the two TTI curves propagated from the independent HTI temperatures have a small systematic offset of about 1~meV/atom over the temperature range considered, with maximum absolute differences below 1.5~meV/atom. The two anchor choices are therefore mutually consistent at this level.

For the melt, Fig.~\ref{fig:si_silica_melt_tti_anchor_consistency} shows that the two TTI curves agree within the statistical uncertainties across the plotted temperature range. The anchor-to-anchor difference remains below 0.4~meV/atom.